\documentclass[twocolumn]{aastex631}

\usepackage{amsmath}
\usepackage{bm}   % FIXME some mathrm vs bm inconsistencies

\begin{document}

\title{Field-Level Inference from Galaxies: BAO Reconstruction}

\author{Adrian E. Bayer} %[0000-0002-3568-3900]
\affiliation{Center for Computational Astrophysics, Flatiron Institute, 162 5th Avenue, New York, NY, 10010, USA}
\affiliation{Department of Astrophysical Sciences, Princeton University, Peyton Hall, Princeton NJ 08544, USA}
\email{abayer@flatironinstitute.org}

\author{Liam Parker} %[0009-0007-4952-1674]
\affiliation{Berkeley Center for Cosmological Physics, Department of Physics, University of California, Berkeley, CA 94720, USA}
\affiliation{Lawrence Berkeley National Lab, 1 Cyclotron Road, Berkeley, CA 94720, USA}

\author{David Valcin} %[0000-0003-0129-0620]
\affiliation{Berkeley Center for Cosmological Physics, Department of Physics, University of California, Berkeley, CA 94720, USA}
\affiliation{Lawrence Berkeley National Lab, 1 Cyclotron Road, Berkeley, CA 94720, USA}

\author{Shi-Fan Chen} %[0000-0002-5762-6405]
\affiliation{Department of Physics, Columbia University, New York, NY 10027, USA}
\affiliation{NASA Hubble Fellowship Program, Einstein Fellow}

\author{Chirag Modi} %[0000-0002-1670-2248]
\affiliation{Center for Cosmology and Particle Physics, New York University, New York, NY 10003, USA}

\author{Uroš Seljak} %[0000-0003-2262-356X]
\affiliation{Berkeley Center for Cosmological Physics, Department of Physics, University of California, Berkeley, CA 94720, USA}
\affiliation{Lawrence Berkeley National Lab, 1 Cyclotron Road, Berkeley, CA 94720, USA}

\begin{abstract}
Baryon acoustic oscillations (BAO) underpin the key cosmological results from modern spectroscopic galaxy surveys, but nonlinear gravitational evolution limits the precision achievable with traditional analysis methods.
To overcome this, we develop field-level inference for BAO, first reconstructing the initial linear density field and then fitting the BAO signal therein.
We benchmark three reconstruction methods: (i) traditional reconstruction based on the Zel'dovich approximation, (ii) explicit field-level inference using differentiable forward modeling with hybrid effective field theory, and (iii) implicit field-level inference using a convolutional neural network to augment traditional reconstruction. 
Using DESI-like Luminous Red Galaxy (LRG) and Bright Galaxy Survey (BGS) catalogs, we find that field-level approaches significantly sharpen the BAO feature relative to traditional reconstruction. 
For LRGs, explicit field-level inference improves constraints on the BAO scale parameters ($\alpha_{\rm iso}, \alpha_{\rm ap}$) by 26\%, while implicit inference improves constraints by 35\%, corresponding to a 2.4$\times$ improvement in figure of merit.
For the higher-density, lower-redshift BGS sample, field-level inference enables information extraction from smaller scales, yielding an improvement in constraints of up to 46\%, corresponding to a 3.2$\times$ improvement in figure of merit.
Crucially, we address longstanding concerns regarding the robustness of field-level reconstruction by leveraging 1,000 mock realizations to perform extensive coverage tests. 
Our results are both unbiased and statistically well-calibrated, maintaining nominal coverage even when using tight simulation-informed priors and under model misspecification.
%Our results establish field-level reconstruction as a validated, high-precision tool capable of maximizing the scientific return of spectroscopic surveys.
\end{abstract}

\section{Introduction} \label{sec:intro}

%Surveys.

%BAO as robust methods to extract info.

%Current methods.

%Another option is field-level. Explain field level and it's potential.

%Here, we measure the BAO in the reconstructed field for the furst time comparing it to traditiaonl methods using the official DESI pipeline.

%We apply to DESI-like galaxies -- a big step up from the typical matter field or halo field field-level inference has aleready been applies to.

Understanding the large-scale structure of the Universe is one of the key objectives of current and upcoming galaxy surveys such as DESI \citep{collaboration2016desi}, PFS \citep{Takada_2014},
Rubin Observatory LSST \citep{LSSTSci},
Euclid \citep{Euclid}, 
SPHEREx \citep{SphereX_2014}, 
SKA \citep{SKA_2009}, and Roman Space Telescope \citep{spergel2013widefield}. These surveys aim to map the three-dimensional distribution of galaxies across cosmic time, providing an unprecedented high-resolution window into the physics of the early Universe, the growth of structure, and the nature of dark energy.

%Understanding the large-scale structure of the Universe is one of the key objectives of current and upcoming galaxy surveys such as DESI\footnote{\url{https://www.desi.lbl.gov}} \citep{collaboration2016desi}, PFS\footnote{\url{https://pfs.ipmu.jp/index.html}} \citep{Takada_2014}, Rubin Observatory LSST\footnote{\url{https://www.lsst.org}} \citep{LSSTSci}, Euclid\footnote{\url{https://www.euclid-ec.org}} \citep{Euclid},  SPHEREx\footnote{\url{https://www.jpl.nasa.gov/missions/spherex}} \citep{SphereX_2014},  SKA\footnote{\url{https://www.skatelescope.org}} \citep{SKA_2009}, and Roman Space Telescope\footnote{\url{https://roman.gsfc.nasa.gov}} \citep{spergel2013widefield}. These surveys aim to map the three-dimensional distribution of galaxies across cosmic time, providing an unprecedented window into the physics of the early Universe, the growth of structure, and the nature of dark energy.

A cornerstone of modern galaxy-survey cosmology is the measurement of baryon acoustic oscillations (BAO), a robust geometric standard ruler imprinted in the matter distribution. Physically, BAO are the fossil imprint of oscillations between baryons and photons in the early Universe  \citep{Peebles70,Sunyaev70}. As the Universe cooled, photons decoupled from baryons, freezing the scale of these oscillations at the sound horizon at drag epoch $r_d$. The characteristics of these oscillations can be seen in various late-time probes, including the galaxy distribution. In the power spectrum, BAO appear as a distinct wiggle pattern superimposed on the smooth broadband component. These wiggles encode a wealth of pristine cosmological information: their amplitude and frequency are governed by the baryon and total matter densities ($\omega_b, \omega_m$), while their phase and shape are sensitive to exotic physics such as free-streaming light species \citep{Bashinsky04, Montefalcone:2025ibh} and primordial isocurvature modes \citep{Zunckel11}. Crucially, isolating the BAO allows one to bypass the modeling uncertainties of the broadband signal; while the broadband shape is subject to degeneracies with galaxy formation physics on nonlinear scales, and there is no universal rule of thumb for the appropriate scale cuts, the BAO feature remains exceptionally robust to these systematics \citep{Angulo2014BAOAstrophysics, Springel2018IllustrisTNGClustering, HernandezAguayo2023MillenniumTNG}.

By measuring the BAO scale, galaxy surveys are able to tightly constrain cosmological parameters. This effort began with the initial detections of BAO in SDSS and 2dFGRS \citep{Eisenstein2005, Cole2005}, and has continued with increasing precision through 6dFGS \citep{Beutler2011}, WiggleZ \citep{Blake2011}, BOSS \citep{Alam2017}, and eBOSS \citep{Alam2021}, extending to photometric surveys like DES \citep{Abbott2022}, high-redshift probes using the Lyman-$\alpha$ forest \citep{Slosar2011, duMasdesBourboux2020}, and most recently with DESI’s second data release \citep{DESI:DR2}.

%Prior to recombination, the high density of the primordial plasma ensured rapid scattering between baryons and photons. In this tight-coupling regime, the two species behaved as a single, nearly perfect fluid. At the level of linear theory, the corresponding Euler and continuity equations reduce to a driven harmonic oscillator with a slowly evolving natural frequency \citep{Peebles70,Sunyaev70}. Overdense regions launched spherical pressure waves in the coupled baryon--photon fluid; as these waves propagated outward, baryon perturbations produced acoustic oscillations that were effectively frozen in once photons decoupled from the baryons. This decoupling sets the characteristic scale of the oscillations---the sound horizon at the drag epoch, $r_d$---after which matter perturbations grow. The observed post-recombination power spectrum can thus be viewed as the projection of these oscillatory density and velocity fluctuations onto the growing mode.

While one can directly measure the BAO feature in the galaxy correlation function or power spectrum, BAO reconstruction is a key tool for making the measurement more precise. Reconstruction seeks to undo the nonlinear large-scale displacements that occurred during the Universe's evolution since the drag epoch, which damp the oscillations in the power spectrum---or, equivalently, smear the BAO peak. By estimating and partially reversing these bulk flows, reconstruction sharpens the BAO feature and improves its statistical precision. Traditional reconstruction is performed using the Zel'dovich approximation, relying on linear theory and discarding information on mildly nonlinear scales, such as non-perturbative BAO damping \citep{Cabass:2023nyo}.

To improve upon this, field-level inference provides an opportunity for optimal measurement of the BAO, by solving the inverse problem mapping from late-time observables to primordial fluctuations and, in turn, reconstructing the full initial density field of the Universe. 
%In the explicit field-level 
%inference approach this is done by 
%Bayesian analysis using the forward model, the likelihood and the prior which depends on 
%cosmology parameters. An alternative is the 
%so called implicit field-level inference 
%which aims to reconstruct initial conditions using Machine Learning Tools. 
%This methodology can in principle extract the maximal amount of cosmological information by jointly fitting initial conditions and cosmology parameters. 
A key goal of field-level inference is to obtain optimal constraints on cosmological parameters by performing inference over all cosmological degrees of freedom, including the initial conditions of the Universe. 
%After marginalizing over the initial conditions, this provides the optimal cosmological constraints. 
While some works suggest large gains \citep{Nguyen:2024yth, SimBIG:2023ywd} compared to traditional summaries---such as the power spectrum and bispectrum---others suggest incremental improvements \citep{Spezzati:2025zsb, Akitsu:2025boy}, highlighting robustness challenges when modeling cosmological data at the field level \citep{bayer2025robust} and performing simulation-based inference on summary statistics \citep{Modi:2023llw}. 

Motivated by the robustness of the BAO to complex small scale physics, here we utilize field-level inference specifically to reconstruct the BAO. 
We use the term \textit{field-level reconstruction} to indicate that we are specifically interested in reconstructing the initial linear field using field-level inference, and subsequently measuring the BAO feature in the power spectrum or correlation function of the reconstructed field.

We consider two classes of field-level inference methods: (1) explicit 
field-level inference, using Bayesian inference with a differentiable forward model of the galaxy field based on hybrid effective field theory (HEFT), and (2) implicit 
field-level inference, using a convolutional neural network (CNN) trained to invert the mapping from the observed galaxy field to the initial density. The former is referred to as \textit{explicit field-level inference} since it involves \textit{explicitly} assuming a form for the posterior, using methods such as optimization \citep{Seljak_2017, Bayer:2022vid, doeser2025learning} or Markov chain Monte Carlo (MCMC) sampling \citep{Jasche_2013, Jasche:2018oym, Schmidt:2018bkr, Schmidt:2020viy, Nguyen:2020hxe, Kostic:2022vok, Bayer:2023rmj, Nguyen:2024yth, Euclid:2024ris, doeser2024bayesian, Simon:2025gwa} for inference. The latter is referred to as \textit{implicit field-level inference} since the posterior, or mean posterior, is \textit{implicitly} learned using a neural network \citep{shallue2023reconstructing, chen2023effective, floss2024improving, Bottema:2025vww, legin2024posterior, cuesta2024joint, Parker:2025mtg}. In this work we use the annealed optimization approach of \cite{Bayer:2022vid} for explicit inference and the architecture of \cite{Parker:2025mtg} for implicit inference. 

A valuable alternative approach of explicit field-level inference for BAO is to jointly MCMC sample the BAO scale and the initial conditions \citep{Babic:2022dws, Babic:2024wph}. However, here we choose to utilize field-level inference strictly as a reconstruction technique, producing a restored density field on which the BAO is subsequently measured from a
Gaussian likelihood analysis of the power spectrum. Our modular ``reconstruct-then-fit'' approach allows for a direct comparison with the standard reconstruction pipeline used by surveys like DESI, and we are also able to perform robust inference  using considerably smaller scales.
%Since the 
%reconstructed field is Gaussian, and the power spectrum contains all the 
%information, 
%our approach should be equivalent to the  
%joint sampling approach. 

To date, field-level reconstruction methods have primarily been tested in idealized settings, such as the dark matter or halo field, which \cite{Parker:2025mtg} showed can yield optimistic results by directly comparing the reconstruction quality for galaxies to their underlying halos. Furthermore, current analyses lack robustness and statistical calibration tests. We thus simulate 1{,}000 DESI-like luminous red galaxy (LRG) and bright galaxy survey (BGS) catalogs. We measure the BAO feature in the reconstructed initial conditions of these mocks and compare to traditional analyses using the same approach as the official DESI BAO pipeline.
We demonstrate that a field-level BAO 
analysis significantly outperforms traditional methods and, crucially, that it is both robust and statistically well-calibrated.

The paper is organized as follows. In Section \ref{sec:method}, we describe the generation of DESI-like LRG and BGS mock catalogs, detail the three reconstruction algorithms (traditional, explicit field-level, and implicit field-level), and outline the BAO fitting methodology. Section \ref{sec:results} presents our main results: studying the reconstructions, analyzing the field-level covariance, quantifying the improvement in BAO constraints, exploring the information content of small scales, and validating robustness against model misspecification. We then conclude in Section \ref{sec:conclusions}. See Figure \ref{fig:summary} and Table \ref{tab:summary} for a summary of our key results. 

\section{Method} \label{sec:method}

In this section, we outline the methodology used to benchmark field-level BAO reconstruction. We begin by describing the generation of DESI-like mock catalogs for LRG and BGS tracers in Section \ref{sec:mocks}. We then detail the three distinct reconstruction algorithms: traditional reconstruction (Section \ref{sec:recon-trad}), explicit field-level inference via differentiable forward modeling (Section \ref{sec:explicit}), and implicit field-level inference via a convolutional neural network (Section \ref{sec:implicit}). We then describe the procedure for extracting the BAO scale from the reconstructed fields (Section \ref{sec:fitting}) and our strategy for testing robustness against model misspecification (Section \ref{sec:robust}).

\subsection{Mocks} \label{sec:mocks}

We generate 1{,}000 LRG and BGS mocks using \texttt{FastPM} \citep{Feng2016,Bayer_2021_fastpm} with $1\,\rm{Gpc}/h$ box length, $1024^3$ particles for LRGs and $2048^3$ for BGS. We use a force grid two times the size of the particle grid. Each mock contains different random phases for the cosmological initial conditions.
We begin the simulation at redshift of $9$ and use 40 steps to evolve to redshift $0.7$ for LRGs and $0.2$ for BGS, approximately corresponding to the effective redshifts of the DESI {LRG2} and {BGS} samples.
We compute the halo catalog using the Friends-of Friends (FoF) algorithm with a linking length of 0.2. 
%and a minimum number of particles per halo of 10.
We use the following cosmological parameters: $\Omega_m = 0.3175$, 
$\Omega_b=0.049$, $h=0.6711$, $n_s=0.9624$, 
$\sigma_8=0.834$, and $M_\nu=0$. We additionally run a set of simulations with $\Omega_m=0.29$ to test robustness to model misspecification.

We apply the \cite{Zheng2005} halo occupation distribution (HOD) model to populate the halos with galaxies. In this framework, the presence of central galaxies in halos is modeled as a Bernoulli process, where the mean occupation number is given by
\begin{equation}
    \langle N_{\text{cent}}^{\text{LRG}}\rangle = \frac{1}{2} \left[1 + \text{erf}\left(\frac{\log M_{\text{h}} - \log M_{\text{cut}}}{\sqrt{2} \sigma_{\log M_{\text{h}}}}\right)\right],
\end{equation}
with $M_{\text{h}}$ representing the halo mass, $M_{\text{cut}}$ the characteristic mass where halos have a 50\% probability of hosting a central galaxy, and $\text{erf}$ the standard error function:
\begin{align}
\text{erf}(x) = \frac{2}{\sqrt{\pi}} \int_0^x e^{-t^2} \, dt.
\end{align}
Satellite galaxies, in contrast, are assumed to follow a Poisson distribution, with a mean occupation number described by a power-law,
\begin{align}
    \langle N_{\text{sat}}^{\text{LRG}}\rangle = \left(\frac{M_{\text{h}} - \kappa M_{\text{cut}}}{M_1}\right)^{\alpha}, \label{eq:sate_hod}
\end{align}
where $\kappa M_{\text{cut}}$ sets the minimum halo mass threshold for hosting satellites, $M_1$ denotes the scale at which a halo hosts on average one satellite, and $\alpha$ controls the steepness of the relation. The spatial distribution of satellites is assumed to follow the Navarro-Frenk-White (NFW) profile~\citep{NFW1996}. Satellite velocities are drawn from a Gaussian centered on the halo velocity, with dispersion matched to that of the dark matter particles within the halo.
Galaxies are shifted into redshift space, and the galaxy field is computed using the cloud-in-cell (CIC) method with \texttt{nbodykit} \citep{Hand_2018}.

To better model the redshift-space clustering of galaxies, we introduce a free velocity dispersion scaling parameter, $v_{\rm disp}$, following the approach of \cite{Variu2023, DESI_fastpm_gal:2024zgo}. This parameter modifies the line-of-sight velocity of satellites as
\begin{align}
    v^{\text{sat, modified}}_{\parallel} = (v^{\text{sat, default}}_{\parallel} - v^{\text{halo}}_{\parallel}) \times v_{\rm disp} + v^{\text{halo}}_{\parallel},
\end{align}
where the subscript $\parallel$ denotes the velocity component along the line of sight.

\begin{figure*}[ht!]
\includegraphics[width=0.5\textwidth]{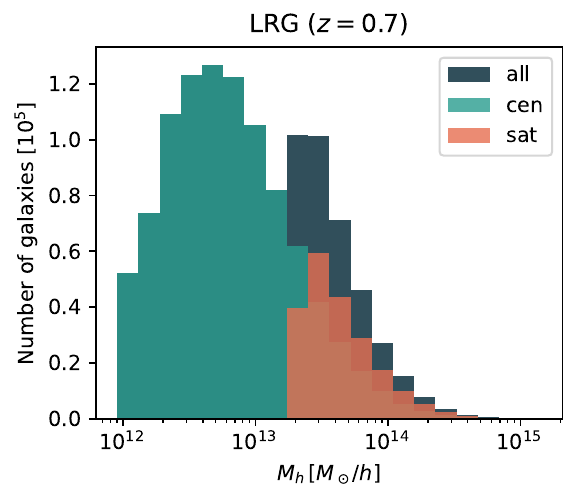}
\includegraphics[width=0.5\textwidth]{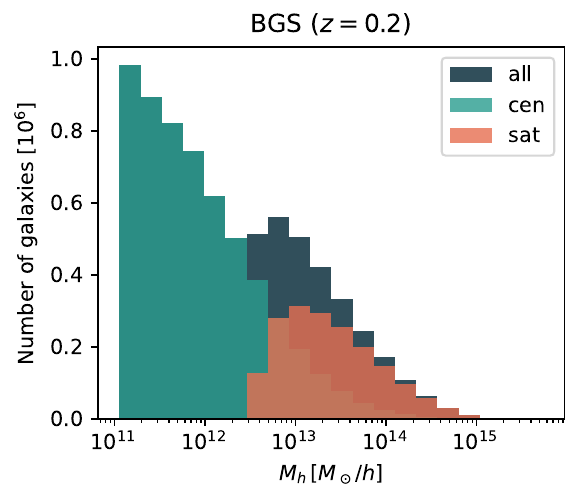}
\caption{
\textbf{Distribution of mock galaxies.}
Number of central (teal), satellite (peach), and all (gray) galaxies as a function of host halo mass for LRGs (left) and BGS (right). 
} 
\label{fig:gal_hist}
\end{figure*}

We use the HOD parameters fit by \cite{DESI_fastpm_gal:2024zgo}, which are calibrated for FastPM, to produce DESI-like mocks. The HOD parameters are shown in Table \ref{tab:sim_params}.
%LRG hodparams = {'alpha':0.12, 'logMmin':12.69, 'logM1':11.71, 'kappa':4.08, 'sigma_logM':0.44}. hod_vdisp = 1.01
%BGS hodparams = {'alpha':0.5271126056024795, 'logMmin':12.262914338999256, 'logM1':12.482083314817826, 'kappa':1.777786673148387, 'sigma_logM':0.9372958876863078}  hod_vdisp = 1.091163297164092
Figure~\ref{fig:gal_hist} shows a histogram of the galaxies produced in one of the mock realizations as a function of their host halo mass for both LRGs and BGS.
%, where the mass of a galaxy $M_g$ is defined as the mass of the host halo divided by the number of galaxies hosted by the host halo. 
In this realization for LRGs, the number density of centrals is $9.7\times10^{-4}\, (h/ {\rm Mpc})^3$, satellites $2.1\times10^{-4}\, (h/ {\rm Mpc})^3$, and the total number density is $1.2\times10^{-3}\, (h/ {\rm Mpc})^3$, while for BGS, the number density of centrals is $5.7\times10^{-3}\, (h/ {\rm Mpc})^3$, satellites $1.8\times10^{-3}\, (h/ {\rm Mpc})^3$, and the total number density is $7.5\times10^{-3}\, (h/ {\rm Mpc})^3$. BGS reaches an order of magnitude lower minimum halo mass, and also has a 6 times larger number density, meaning it has a higher signal-to-noise on small scales and thus a potential for greater improvement from a field-level approach.
Note that, until now, DESI uses a subsample of BGS for its BAO analysis, finding that this is sufficient for optimal traditional BAO reconstruction \citep{KP4, Krolewski2025DESIDR1SpecSys}, however, we will benchmark field-level approaches using the full BGS sample in this work. Table \ref{tab:sim_params} summarizes the simulation details.

\begin{table}[t]
    \centering
    \begin{tabular}{l||c|c}
        \hline
        Parameter & LRG & BGS \\
        \hline
        Num.~Realizations & 1000 & 1000 \\
        Box Length $[\mathrm{Gpc}/h]$ & 1 & 1 \\
        Particle Grid & $1024^3$ & $2048^3$ \\
        Force Grid & $2048^3$ & $4096^3$ \\
        Snapshot Redshift & $0.7$ & $0.2$ \\
        Number Density $[(h/\mathrm{Mpc})^{3}]$ & $1.2 \times 10^{-3}$ & $7.5 \times 10^{-3}$ \\
        \hline
        $\alpha$ & 0.12 & 0.53 \\
        $\log M_{\rm cut}$ & 12.69 & 12.26 \\
        $\log M_{1}$ & 11.71 & 12.48 \\
        $\kappa$ & 4.08 & 1.78 \\
        $\sigma_{\log M_h}$ & 0.44 & 0.94 \\
        $\,v_{\mathrm{disp}}$ & 1.01 & 1.09 \\
        \hline
    \end{tabular}
    \caption{Summary of simulation parameters and galaxy sample properties for the LRG and BGS mocks.}
    \label{tab:sim_params}
\end{table}

%, %alo 3.1e-03 (h/Mpc)^3  % log M_0 = \log(4.08)+12.69

% minimum halo mass of $M_{\rm min}\simeq ?? M_\odot / h$, and a bias of $b_1\simeq4.0$.

\subsection{Reconstruction}

\subsubsection{Traditional BAO Reconstruction}
\label{sec:recon-trad}

Traditional BAO reconstruction aims to reverse the nonlinear gravitational evolution of the density field, sharpening the acoustic peak and partially restoring the linear correlation between observed structure and the initial conditions. The key idea is to estimate and subtract the Lagrangian displacement field, $\boldsymbol{\Psi}(\textbf{q}, t)$, which maps a particle’s initial Lagrangian coordinate $\textbf{q}$ to its Eulerian position at time $t$:
\begin{align}
\textbf{x}(\textbf{q},t) = \textbf{q} + \boldsymbol{\Psi}(\textbf{q},t).
\end{align}
By estimating these displacements from the observed galaxy field and then shifting the galaxies backwards, one effectively rewinds the formation of structure, yielding a field closer to the linear regime.

The displacement field $\boldsymbol{\Psi}$ is computed using the Zel’dovich approximation \citep{zel1970gravitational}, which models the displacement as the gradient of a potential. In real space, the displacement field satisfies the relation:
\begin{align}
    \nabla \cdot \boldsymbol{\Psi} = - \frac{\delta}{b},
\end{align}
where $\delta$ is the smoothed overdensity field and $b$ is the linear bias relating the tracer field to the matter distribution. In redshift space, distortions along the line of sight modify this relation, introducing an additional divergence term:
\begin{equation}
    \nabla \cdot \boldsymbol{\Psi} + f\nabla \cdot \left[(\boldsymbol{\Psi} \cdot \hat r) \hat r\right] = - \frac{\delta}{b},
\label{eq:rsd_lag}
\end{equation}
where $f$ is the linear growth rate and $\hat r$ is the unit vector along the line of sight.

To solve for the displacement field, we use the iterative Fourier-based method, \textsc{IterativeFFT} \citep{burden2015reconstruction}, which 
%is significantly faster than configuration-space approaches and accurate at moderate grid resolution. This 
is the approach adopted by DESI. This method assumes that $\boldsymbol{\Psi}$ is irrotational, allowing it to be expressed as the gradient of a scalar potential. However, the redshift-space term $(\boldsymbol{\Psi} \cdot \hat r)\hat r$ is not irrotational, which complicates the solution of \eqref{eq:rsd_lag} in Fourier space.
To address this, the redshift-space contribution is decomposed into irrotational and solenoidal components and solve iteratively using a plane-parallel approximation ($\hat r \rightarrow \hat x$). This yields the following expression for the displacement field after $n$ iterations:
\begin{equation}
    \boldsymbol{\Psi}_{\textrm{FFT}, n} = - \frac{i \textbf{k} \delta_{\textrm{g,real}}}{bk^2} \left[1 + (-f)^{n+1} \left(\frac{k_x}{k}\right)^{2(n+1)} \right],
\label{eq:ifft}
\end{equation}
where $\delta_{\textrm{g,real}}$ is the smoothed real-space galaxy overdensity field.

Methodologically, we begin by smoothing the input tracer field using a Gaussian kernel with width $15\,{\rm Mpc}/h$ to suppress small-scale nonlinearities, as this is the value used by DESI \citep{DESI:DR2, Chen:2024eri}. 
%We also explore the dependence on the smoothing scale. 
We also generate $20N_g$ uniformly distributed random particles (where $N_g$ is the number of galaxies) to sample the survey volume. We then solve \eqref{eq:ifft} iteratively to obtain $\boldsymbol{\Psi}$ for both tracers and randoms.
The bias $b$ is estimated by fitting the power spectrum of the tracer field to that of a biased linear matter field:
\begin{align}
    b = \arg\min_b \sum_{|\mathbf{k}| \in (0, 0.1)} \left[P_\mathrm{g}(\mathbf{k}) - b^2 P_\mathrm{lin}(\mathbf{k})\right]^2,
\end{align}
where $P_\mathrm{g}(k)$ is the power spectrum of the galaxy field, and $P_\mathrm{lin}(k)$ is the linear theory prediction. A fiducial value of $f$ is chosen for the reconstruction, and we investigate the robustness to this choice.

After computing $\boldsymbol{\Psi}$, we shift both tracers and randoms by $-\boldsymbol{\Psi}(\textbf{x})$, and deposit them onto separate grids using the Cloud-in-Cell (CIC) assignment scheme \citep{hockney2021computer}. We shift the randoms including the RSD displacements, following the \texttt{recsym} convention, which is more robust than \texttt{reciso} \citep{Chen:2024tfp}. The final reconstructed field is obtained by subtracting the displaced randoms from the displaced tracers and dividing by the estimated bias:
\begin{align} 
\delta_\mathrm{rec}(\mathbf{x}) = \frac{\hat \delta_\mathrm{g}(\mathbf{x}) - \hat \delta_\mathrm{r}(\mathbf{x})}{b},
\end{align}
where $\hat \delta_\mathrm{g}(\mathbf{x})$ and $\hat \delta_\mathrm{r}(\mathbf{x})$ are the CIC-deposited grids of displaced tracers and randoms, respectively.

We implement traditional BAO reconstruction using the \texttt{pyrecon}\footnote{\url{https://github.com/cosmodesi/pyrecon}} package.

\subsubsection{Explicit Field-Level Inference}
\label{sec:explicit}

In this approach, Bayesian inference is performed using differentiable forward modeling. The posterior of the linear density field $\delta_1$ is explicitly written down and optimized. Given an observation of the galaxy field $\delta_g$, the posterior for the underlying linear field $\delta_1$ can be written (up to a constant) as 
\begin{align} \label{eq:field_posterior}
    -2\log p&(\delta_1 | \delta_g) \nonumber \\
    &= \sum_{\bm{k}} \left[ \frac{|\delta_g(\bm{k}) - f_g(\bm{k};\delta_1)|^2}{\sigma_g^2(\bm{k})} + \frac{|\delta_1(\bm{k})|^2}{P_1(k)} \right],
\end{align}
where $f_g$ is the forward model of the galaxy field, $\sigma_g$ is the error in the forward model of the galaxy field, and $P_1$ is the linear power spectrum with BAO wiggles removed (this is done for robustness, and is explained below). The sum is performed over all Fourier-space pixels. The first term represents a Gaussian likelihood term, while the second term is the prior representing the linear field is a Gaussian random field with variance given by the power spectrum. Note that the density field is evaluated in Fourier space, as the model error is $k$ dependent (as discussed below). Treating the EFT-based likelihood as Gaussian is only correct on linear scales before non-Gaussian noise sets in \citep{Akitsu:2025boy}, thus a more sophisticated likelihood---or implicit inference---is required to accurately reconstruct small scales. Nevertheless, employing a Gaussian likelihood for explicit inference is currently state-of-the-art and produces high quality reconstruction. As in the traditional approach for BAO reconstruction, a fiducial cosmology is assumed throughout the reconstruction process.

The forward model $f_g$ is computed using HEFT \citep{Modi:2019qbt}, which is the most accurate differentiable forward model of galaxy clustering currently available (see e.g.~\cite{jfof} and \cite{diffhod} for efforts to move beyond HEFT by making halo and galaxy identification differentiable). We first compute the matter overdensity field and displacements $\psi$ using \texttt{pmwd} \citep{li2022pmwd} (a \texttt{jax} version of \texttt{FastPM}). We use 5 steps between redshift $9$ and the sample redshift. We use $N_{\rm c} = 256^3$ matter particles, and force grid with resolution $N_{\rm grid} = 256^3$, limited by the memory constraints of a single GPU. For the matter-galaxy connection we use second-order HEFT, such that 
\begin{align} \label{eqn:bias}
    f_g(\bm{k}) = \int d^3\bm{q}\,\Big[& 1 + b_1 \delta_1(\bm{q})
    + b_2 \left( \delta_1^2(\bm{q}) - \sigma_1^2 \right)
    \nonumber\\
    &+ b_{s^2}\left(s^2(\bm{q})-\langle s^2\rangle\right)
    \nonumber\\
    &+ b_{\nabla^2}\,\nabla^2\delta_1(\bm{q})
    \nonumber\\
    &+ b_{\mu^2\nabla^2}\,\mu^2\,\nabla^2\delta_1(\bm{q})
    \nonumber\\
    &+ b_{\mu^4\nabla^2}\,\mu^4\,\nabla^2\delta_1(\bm{q})
    \Big]\,
    e^{ - i\bm{k} \cdot \left( \bm{q} + \bm{\psi}(\bm{q}) \right)},
\end{align}
where $\bm{k}$ is the wavevector, $\bm{q}$ is the Lagrangian grid coordinates, $\delta_1(\bm{q})$ is the linear overdensity, $\sigma_1^2\equiv\langle\delta_1^2\rangle$ is its variance, and $\bm{\psi}(\bm{q})$ is the Lagrangian displacement field computed using \texttt{pmwd}. The operator $s^2(\bm{q})\equiv s_{ij}(\bm{q})s_{ij}(\bm{q})$ is the tidal shear invariant with $s_{ij}(\bm{q})\equiv\left(\partial_i\partial_j/\nabla^2-\delta_{ij}/3\right)\delta_1(\bm{q})$. The derivative operator $\nabla^2\delta_1$ and its $\mu$-dependent extensions capture leading EFT RSD corrections; in Fourier space they correspond to factors $-k^2\delta_1(\bm{k})$, $-k^2\mu^2\delta_1(\bm{k})$, and $-k^2\mu^4\delta_1(\bm{k})$ (see e.g.,~\cite{Schmittfull:2020trd, Stadler:2023hea, Stadler:2024aff})
%We find that the shear and $\nabla^2$ terms do not improve the reconstruction, and thus drop them from the analysis in the main paper, but show results in Appendix \ref{app:b2plus}. %--- but see \cite{Nguyen_2021} for a detailed review of cases where incorrect bias modeling can bias inference (for HMC, but here we are optimizing).
%
The error $\sigma_g$ corresponds to the error power spectrum,
\begin{equation}
    \sigma^2_g(k) = P_{\rm err} (k) = \frac{1}{N_{\rm modes}(k)} \sum_{\bm{k}:|\bm{k}|=k} |\epsilon(\bm{k})|^2,
\end{equation}
where $N_{\rm modes}(k)$ in the number of modes in the $k$ bin, and $\epsilon (\bm{k}) \equiv \delta_g(\bm{k}) - f_g(\bm{k})$ is the error in the forward model in describing the observed data. The error quantifies the inaccuracy of HEFT in modeling a galaxy sample generated by HOD. Based on EFT, the error variance can be parameterized to second order as follows
\begin{equation}
    \sigma_g^2 = A+Bk^2+C\mu^2 +Dk^2\mu^2,
\end{equation}
where $A,B,C$, and $D$ are free parameters.
We fix these parameters and the bias parameters by minimizing $\sigma_g$ given $\delta_g$ before performing reconstruction. %In a more complete analysis one could jointly sample these free parameters during reconstruction. -- conc

Note we find in Section~\ref{sec:results} (Figure~\ref{fig:summary}) that excluding the $s^2$ and $\nabla^2$ terms from the forward model produces identical reconstructions and BAO constraints, suggesting that the leading-order displacement and density bias are sufficient for BAO reconstruction in our reconstruct-then-fit approach, and that higher-order bias terms are unlikely to yield further improvement. We thus drop these terms in all but the fiducial LRG case.

For the linear power in the prior $P_1$ we remove wiggles using the approach of \cite{Vlah:2015zda}. As the forward model and likelihood are not accurate to small scales, the inferred $\delta_1$ is influenced by the prior. If the prior contains BAO wiggles corresponding to a particular fiducial cosmology, it will simply learn the prior, which in the case of cosmological model misspecification will lead to biased or overconfident inference. This bias could be reduced by marginalizing over cosmological parameters while fitting the BAO, however, in this work we consider only an optimization-based point estimator of the initial conditions---mimicking the traditional BAO reconstruction approach---and thus remove the BAO wiggles from the prior to ensure robustness.

We maximize the posterior to get the maximum a posteriori (MAP) estimate of the initial field. 
Since the parameter space consists of $256^3 \approx 1.6\times10^7$ dimensions, we use a gradient-based optimization algorithm to ensure tractable inference, requiring a differentiable simulator such as \texttt{pmwd}. In this work, we use the LBFGS-B algorithm \citep{Byrd_1995:LBFGS} which uses gradients at each step, and additionally keeps track of them over the trajectory to approximate the Hessian with a low memory cost. 
As there are many more modes to be fitted on small scales than large scales, we anneal the posterior to gradually learn modes starting from large scales and transitioning to small scales by iteratively fitting modes up to a given scale $k<k_{\rm iter}$ and gradually increasing $k_{\rm iter}$, as in \cite{Bayer:2022vid}, to ensure the many low signal-to-noise small scale modes do not prevent learning from the few high signal-to-noise large scale modes. 
We perform annealing by gradually decreasing the width of a Gaussian smoothing kernel to  $1000/1024\approx1\,{\rm Mpc}/h$, which we find to be optimal; annealing to $\approx 2 {\rm Mpc}/h$ gives essentially identical performance for BAO inference, while stopping at a higher smoothing scale yields suboptimal constraints.

\subsubsection{Implicit Field-Level Inference}
\label{sec:implicit}

In this approach, the mean of the posterior is learned using machine learning through simulation-based inference (SBI). We follow the approach of \cite{Parker:2025mtg}, using a CNN to perform subgrid corrections on top of the traditional BAO reconstruction (described in Section \ref{sec:recon-trad}) to improve small scale convergence.

A CNN is trained to refine the density field estimate obtained via traditional reconstruction by incorporating additional information from the nonlinear tracer field. This approach is physics-informed, as we do not ask the CNN to learn the complete mapping from the galaxy data to the initial conditions, rather we provide it with the physics of traditional reconstruction and ask it to learn the residual correction, enabling both interpretability and efficiency. Specifically, the network learns a mapping
\begin{equation}
    f: (\delta_{\mathrm{g}}(\mathbf{x}), \delta_{\mathrm{trad}}(\mathbf{x})) \rightarrow \delta_{\mathrm{l}}(\mathbf{x}),
\end{equation}
where $\delta_{\mathrm{g}}$ is the observed galaxy field, and $\delta_{\mathrm{trad}}$ is the traditional reconstruction of $\delta_\mathrm{g}$.
These two fields are input as separate channels into the network, which processes local subgrids of size $N_{\mathrm{sub}}^3 \times 2$. The CNN consists of 9 double-convolutional layers with \textsc{ReLU} activations, where each block includes one padded and one unpadded convolution to reduce spatial dimensions and avoid boundary artifacts. The output is a corrected estimate of the linear initial conditions over a region of size $(N_{\mathrm{sub}} - 18)^3$. The channel width increases with depth: 32 channels in the first three layers, 64 in the next three, and 128 in the final three.
We use $N_{\mathrm{sub}}=50$, corresponding to a subgrid length of $\approx195\,{\rm Mpc}/h$, sufficiently larger than the BAO scale. We average the overlapping predictions to produce the final output field. 

To train the model, a Fourier-space loss function is used:
\begin{align}
    \mathcal{L}_{\mathrm{Fourier}} = \sum_{\textbf{k}} M(k) \left| f(\Tilde{\delta}_{\mathrm{g}}(\textbf{k}), \Tilde{\delta}_{\mathrm{rec}}(\textbf{k})) - \Tilde{\delta}_{\mathrm{l}}(\textbf{k}) \right|^2,
\end{align}
where tildes denote Fourier transforms, and $M(k)$ is a scale-dependent weight:
\begin{align}
M(k) =
\begin{cases} 
10, & \text{if } k \in \left[0.08, 0.5\right]\,h\,\mathrm{Mpc}^{-1} \\
1, & \text{otherwise}.
\end{cases}
\end{align}
This weighting emphasizes the nonlinear regime ($k \gtrsim 0.08\,h\,\mathrm{Mpc}^{-1}$) while still fitting the full spectrum. The maximum $k$ value is to respect the convergence of the simulation. During inference, we apply the trained network across the entire volume by extracting overlapping patches with a stride of $N_{\mathrm{sub}} - 40$, ensuring that each voxel is predicted multiple times in different contexts. This patchwise inference scheme allows us to scale to arbitrarily large cosmological volumes, avoids discontinuities at subgrid boundaries, and is robust to super-sample effects~\citep{Bayer:2022nws}. It also enables training with relatively few large-volume simulations; we use 100 simulations for training, which is sufficient for convergence \citep{Parker:2025mtg}, leaving 900 for evaluation.

\subsection{Fitting the BAO}
\label{sec:fitting}

The BAO scale is used as a standard ruler, corresponding to a peak in the correlation function at separation $r_d$, or equivalently, harmonic oscillations in Fourier space with frequency $2\pi/r_d$. To measure the BAO, spectroscopic surveys convert observed redshifts and positions into comoving coordinates based on a reference \textit{fiducial} cosmology.
Thus, at a given redshift $z$, the parallel and perpendicular to line-of-sight components of the BAO constrain $H(z)r_d$ and $D_A(z)/r_d$ respectively. 
The resulting clustering statistics---such as power spectrum multipoles $P_\ell(k)$ or correlation function multipoles $\xi_\ell(s)$---are then fitted against a theoretical BAO template. This template is allowed to rescale along the parallel and perpendicular directions \citep{Padmanabhan08}, yielding the dilation parameters
\begin{equation}
    \alpha_{\parallel} = \frac{H^{\mathrm{fid}}(z)r^{\mathrm{tem}}_{d}}{H(z)r_{d}}, \qquad \qquad \alpha_{\perp} = \frac{D_{A}(z)r^{\mathrm{tem}}_{d}}{D^{\mathrm{fid}}_{A}(z)r_{d}}.
    \label{eqn:alpha_defs}
\end{equation}
%such that $k$ and $\mu$ are dilated as follows,
%\begin{align}
%    k' &= \frac{k}{\alpha_\perp}\left[1 + \mu^2 \left(\frac{1}{\alpha_{\rm ap}^2} - 1 \right) \right]^{1/2}, \\
%    \mu' &= \frac{\mu}{\alpha_{\rm ap}} \left[1 + \mu^2 \left(\frac{1}{\alpha_{\rm ap}^2} - 1 \right) \right]^{-1/2}\,.
%\end{align}
In these expressions, ``fid'' and ``tem'' denote quantities calculated in the fiducial and template cosmologies, respectively. It is standard practice to re-parameterize these into an isotropic scaling, $\alpha_{\mathrm{iso}}\equiv\alpha_{\parallel}^{1/3}\alpha_{\perp}^{2/3}$, and an anisotropic warping parameter, $\alpha_{\mathrm{ap}} \equiv \alpha_{\parallel}/\alpha_{\perp}$. The tighter one can constrain these parameters, the more precisely one has measured the BAO scale, and in turn the more precise the cosmological constraints obtained from BAO reconstruction.

For all of our reconstruction algorithms, we measure the post-reconstruction power spectrum $P$, or 2-point correlation function $\xi$, and then fit the model of \cite{Chen:2024tfp}:
\begin{equation} \label{eq:generic_model_chen}
    P(k, \mu) = \mathcal{B}(k, \mu) P_{\rm nw}(k) + \mathcal{C}(k, \mu)P_{\rm w}(k) + \mathcal{D}(k)\,,
\end{equation}
where $P_{\rm nw}(k)$ and $P_{\rm w}(k)$ denote the smooth (no-wiggle) and BAO (wiggle) components of the power spectrum. The matter power spectrum template is predicted from \texttt{CLASS} \citep{CLASS}, and the wiggle component is computed using the fitting formula of \cite{Wallisch:2018rzj}.
Following \cite{seo2016foreground, Beutler2017}, $\mathcal{B}(k, \mu)$ is given by 
\begin{equation}
\label{Chen2}
    \mathcal{B}(k,\mu) = \left(b+f\mu^{2}(1 - s(k)\right)^{2} F_{\rm fog}\,,
\end{equation}
where $b$ is the linear bias, $f$ is the growth rate, and $F_{\rm fog} = \left(1 + \frac{1}{2} k^2\mu^{2} \Sigma_s^2\right)^{-2}$ accounts for the `Fingers of God' caused by halo virialization \citep{Park1994}. 
For traditional reconstruction, we use the \texttt{recsym} convention, and thus $s(k) = 0$ \citep{Chen:2024tfp}. For explicit field-level inference we use $s(k) = \exp\left[-(k \Sigma_{\rm sm})^2/2 \right]$, with $\Sigma_{\rm sm}$ as the smallest smoothing scale in the annealing ($1\,{\rm Mpc} /h$) \citep{Chen:2024tfp}. For implicit field level inference we use $s(k)=1$, effectively corresponding to an infinitesimal smoothing scale.
%We use the \texttt{reciso} convention, where $s(k) = \exp\left[-(k \Sigma_{\rm sm})^2/2 \right]$ and $\Sigma_{\rm sm}$ is the reconstruction smoothing scale.
%
$\mathcal{C}(k, \mu)$ captures the anisotropic nonlinear damping on the BAO feature,
\begin{equation}
    \mathcal{C}(k,\mu) = \left(b+f\mu^{2}\right)^{2}\exp\left[-\frac{1}{2}k^2\biggl(\mu^{2}\Sigma_{||}^2 + (1-\mu^{2})\Sigma^{2}_{\perp}\biggl)\right],
    \label{Chen3}
\end{equation}
where $\Sigma_{||}$ and $\Sigma_{\perp}$ capture the damping for modes along and perpendicular to the line of sight. $\mathcal{D}(k)$ captures deviations between the broadband shape of the power spectrum and linear theory. Following \cite{Chen:2024tfp}, we use a spline basis,
\begin{equation} \label{eq:spline_basis}
    \mathcal{D}_\ell(k > k_{\rm min}) = \sum_{n=-1}^7 a_n W_3\left(\frac{k}{\Delta} - n\right)\,,
\end{equation}
where $W_3$ is a piecewise cubic spline kernel \citep{Chaniotis2004, Sefusatti2016:1512.07295}, and the spacing is chosen as $\Delta = 0.06\, h/{\rm Mpc}$ in order to match the broadband shape of the power spectrum without producing spurious BAO wiggles.

Given the Gaussian nature of the linear power spectrum, we fit this model using a Gaussian likelihood to infer the BAO scale parameters $\alpha_{\rm iso}$ and $\alpha_{\rm ap}$. To bin the power spectrum, we use the typical DESI settings \citep{Paillas:2024cru}, going from $k_{\rm min}=0.02\,h/{\rm Mpc}$ to $k_{\rm max}=0.3\,h/{\rm Mpc}$ in steps of $dk=0.005\,h/{\rm Mpc}$; we also investigate the gains of including smaller scales by increasing $k_{\rm max}$. The covariance is computed numerically using the mocks: For each realization, we compute a leave-one-out covariance from the other mocks and apply Ledoit–Wolf regularization to obtain a well-conditioned estimate. 
We also compare to the theoretical Gaussian covariance, which for the monopole is 
\begin{equation} \label{eqn:gauss_cov}
    C^G_{ij} = \delta_{ij} \frac{2}{N_{k_i}} \left[ P(k_i) + P_{\mathrm{noise}}(k_i) \right]^2,
\end{equation}
where $N_{k_i}$ is the number of independent modes in the $i$-th $k$-bin, and $P_{\mathrm{noise}}$ represents the shot-noise or residual noise component. For the field-level reconstruction, which recovers the linear field, we compare simply to the cosmic variance limit where $P_{\mathrm{noise}} \to 0$. 
%We use `$\sigma$' as a shorthand for the Gaussian covariance.

We consider two fitting configurations: marginalizing using uniform priors or fixing the nuisance parameters $(b, f, \Sigma_s, \Sigma_\parallel, \Sigma_\perp, \{ a_n\})$. When marginalizing, we use large uniform priors on all of the nuisance parameters as in \cite{Philcox:2020vvt}. The approach taken by DESI is to use more restrictive, informed, priors on the nuisance parameters \citep{Paillas:2024cru, DESI:DR2}, thus our fixed nuisance parameter experiment provides a more optimistic forecast and tests sensitivity to the choice of prior using coverage tests, which has never been done even in the context of traditional reconstruction. We fix the nuisance parameters by fitting the model, taking the MAP of the nuisance parameters, fixing them, and then fitting the BAO parameters. We sample for the parameters using the \texttt{emcee} sampler \citep{emcee:1202.3665} using the \texttt{desilike}\footnote{\url{https://github.com/cosmodesi/desilike}} package.

To assess the quality of improvement of field-level methods compared to the traditional method, we consider both the percentage improvement of constraint for both $\alpha_{\rm iso}$ and $\alpha_{\rm ap}$, and the increase in figure of merit (FoM), corresponding to the area under curve:
\begin{equation}
    {\rm FoM} = \frac{1}{\sqrt{\det \, {\rm Cov}(\alpha_{\rm iso},\alpha_{\rm ap})}} \simeq \frac{1}{\sigma_{\alpha_{\rm iso}} \sigma_{\alpha_{\rm ap}}}.
\end{equation}

\subsection{Model Misspecification} \label{sec:robust}

As outlined in Section \ref{sec:fitting}, there are three locations where cosmological parameters are assumed. The first is in the reconstruction algorithm---for traditional reconstruction a fiducial $f$ is assumed, for explicit inference a fiducial cosmology is assumed in the forward model and prior, and for implicit inference the neural network training is performed at a fiducial cosmology. The second is in the BAO model template, which affects the sound horizon (the quantity with superscript `tem' in equation (\ref{eqn:alpha_defs})). The third is a geometric effect due to converting distances from observable to comoving coordinates (affecting quantities with superscript `fid' in equation (\ref{eqn:alpha_defs})). An incorrect choice of cosmological parameters for any of these three procedures could bias results. Moreover, in each reconstruction algorithm, the galaxy bias, or HOD parameters (in the case of implicit inference) are also fit or chosen as some fiducial values.
A thorough study of the effect of model misspecification on traditional BAO reconstruction is provided by \cite{cosmology_dependence}, which shows BAO reconstruction to be an extremely robust procedure, partly due to the flexible fitting procedure absorbing much of the misspecification.

We explicitly consider the effect of model misspecification on field-level BAO reconstruction by performing a BAO fit using the same fiducial cosmology as above, but applied to test data with: 
\begin{enumerate}
    \item \textbf{Misspecified Cosmology:} an incorrect value of $\Omega_m=0.29$ is used, with all other parameters the same as before. This modifies $f$, $H(z)$, $r_d$, and the galaxy bias parameters, testing the first two sources of model misspecification. In particular, in the absence of Alcock-Paczynski (AP) distortions \citep{Alcock1979}, this simply causes an isotropic rescaling by $r_d$, giving $\alpha_{\rm iso}\approx0.979$ and $\alpha_{\rm ap}=1$.
    \item \textbf{AP Distortions:} all positions in the $z$ direction are multiplied by a factor of $1+10/256$, extending the size of the field in that direction by 10 pixels. This mimics an incorrect conversion of observable to comoving coordinates, testing the third source of model misspecification. This gives $\alpha_\parallel\approx0.962$, and in turn $\alpha_{\rm iso}\approx0.987$ and $\alpha_{\rm ap}\approx0.962$.
\end{enumerate}

In all test cases, we perform rigorous coverage tests to validate the statistical reliability of our error estimates. While it is standard practice to check for bias in the mean, it is equally critical to verify that the inferred uncertainties accurately reflect the true scatter of the estimator. Performing such tests is often prohibited by the limited number of survey mocks available (for example, DESI utilizes $\sim$25 high-resolution AbacusSummit simulations \citep{Maksimova2021} for validation). Moreover, this is particularly challenging for explicit field-level inference, where the high computational cost of forward modeling and optimization typically restricts validation to just a handful of realizations. In this work, we overcome these barriers and run the full reconstruction and fitting pipeline on 1,000 independent mock realizations.

\section{Results} \label{sec:results}

In this section, we present the performance of the three reconstruction methods. We begin in Section \ref{sec:results-lrg} by analyzing the LRG sample, where we inspect the reconstructed fields visually and statistically, study the covariance matrices, and present the BAO constraints under different nuisance parameter priors. In Section \ref{sec:results-misspec}, we test the robustness of the methods to cosmological model misspecification. In Section \ref{sec:results-bgs}, we extend this analysis to the higher-density BGS sample. We then investigate the information content of small scales in Section \ref{sec:results-kmax} by varying the maximum wavenumber $k_{\rm max}$ used in the fit. Finally, we summarize our findings and compare to results using the correlation function in Section \ref{sec:results-summary}.

\subsection{LRGs} \label{sec:results-lrg}

\begin{figure*}[ht!]
\includegraphics[width=\textwidth]{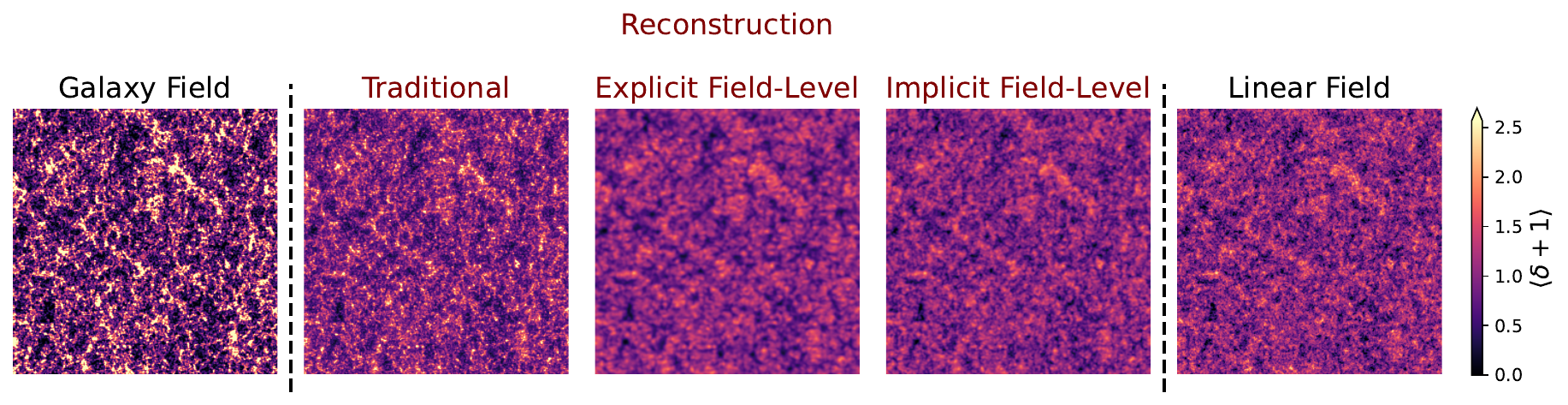}
\caption{
\textbf{Illustration of Reconstructions.} Given the observed galaxy field (left), we perform reconstruction using three different methods of increasing quality from left to right: traditional BAO reconstruction, explicit field-level inference, and implicit field level inference. The reconstruction quality can be seen to improve going from traditional to explicit field-level to implicit field-level. All images are 2D projections along a direction perpendicular to the line-of-sight of thickness $125\,{\rm Mpc}/h$. The colorbar limits are set based on the extreme values of the linear field. The linear bias has been scaled out of the traditional reconstruction to enable apples-to-apples comparison.}
\label{fig:fields}
\end{figure*}

To qualitatively study the reconstruction quality, Figure \ref{fig:fields} shows 2D projections of the reconstructed fields alongside the galaxy field data (left) and the corresponding linear field (right) for a particular data realization. The reconstruction quality can be seen to improve going from traditional to explicit field-level to implicit field-level. Traditional reconstruction has undone much of the nonlinear evolution, but still has many highly overdense and underdense structures due to the inability of the Zel'dovich approximation to capture small scales, and because traditional reconstruction simply shifts the galaxies to sharpen the BAO feature, which does not directly produce the smooth linear field. On the other hand, both field-level approaches directly target the linear field at the map level, with the implicit approach producing a slightly sharper image than the explicit approach due to its ability to more accurately capture small scales.

\begin{figure}[ht!]
\hspace{-1.2em}
\includegraphics[width=0.5\textwidth]{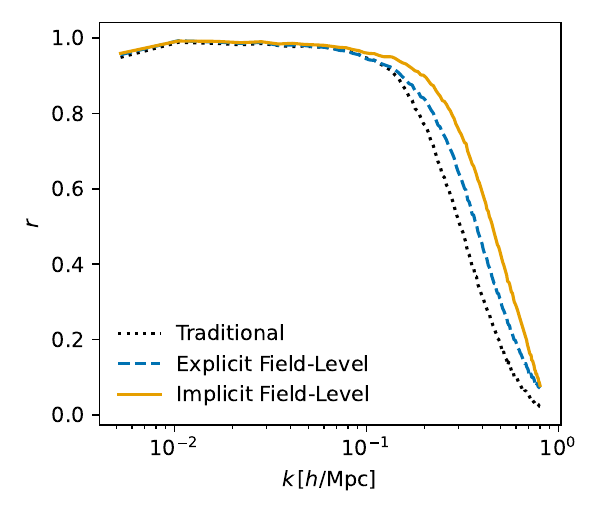}
\caption{
\textbf{Cross-correlation coefficient} $r$ of traditional (black), explicit field-level (blue), and implicit field-level (gold) reconstruction, in order of increasing quality.
}
\label{fig:bao_r}
\end{figure}

To quantitatively study the reconstruction quality, Figure \ref{fig:bao_r} shows the cross-correlation coefficient $r$ of each reconstructed field with the true linear field averaged over 900 realizations. While all methods give similar reconstruction quality on large scales, the small scale reconstruction quality is again seen to be best for implicit field-level, followed by explicit field-level. Traditional performs the worst as the Zel'dovich approximation is unable to capture small scale features. We note that traditional reconstruction is not strictly designed to reconstruct the linear field, only to sharpen the BAO, but it nevertheless is instructive to compare traditional reconstruction to the linear field as $r$ relates to the information of the BAO. Explicit field-level inference improves upon this by using higher-order HEFT to capture further information on small scales, but it too is limited by the accuracy of HEFT in describing the true galaxy data. Implicit field-level inference uncovers optimal reconstruction on the smallest of scales as it is trained on simulations that are in distribution with the data -- the decorrelation on small scales is due to the fundamental shot noise limit.

\begin{figure}[ht!]
\hspace{-1.2em}
\includegraphics[width=0.5\textwidth]{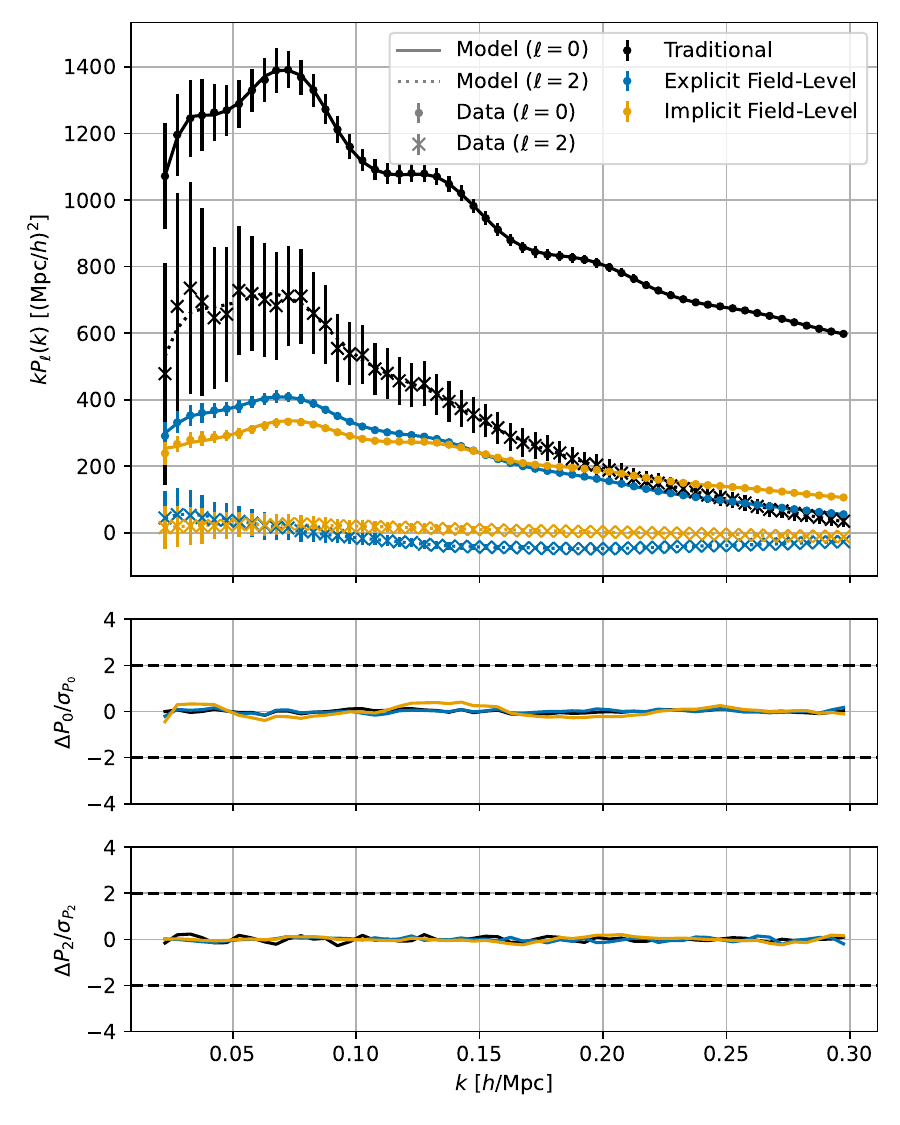}
\caption{
\textbf{Monopole and Quadrupole} of reconstructed fields for traditional (black), explicit field-level (blue), and implicit field-level (gold) reconstruction. The top panel shows the monopole (solid model fit, circular data points) and quadrupole (dotted model fit, crossed data points). The lower panels show the residuals divided by the diagonal covariance, showing sub-percent quality of fit.
}
\label{fig:bao_fit}
\end{figure}

\begin{figure*}[ht!]
\includegraphics[width=\textwidth]{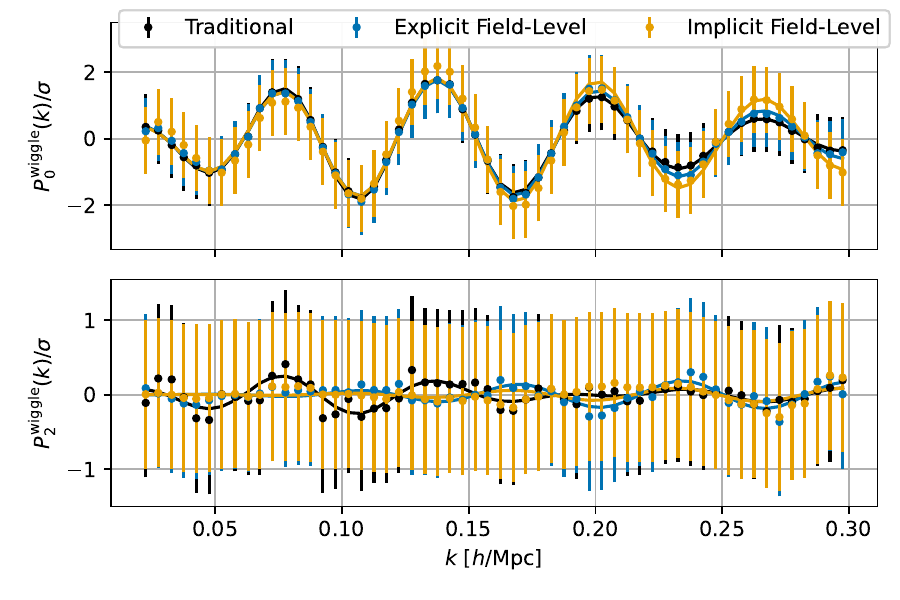}
\caption{
\textbf{BAO wiggle} signal-to-noise of reconstructed fields for traditional (black), explicit field-level (blue), and implicit field-level (gold) reconstruction. Each panel shows the power divided by the root Gaussian covariance. The small scale wiggles in the monopole (top panel) are best reconstructed by implicit inference, followed by explicit. The quadrupole (lower panel) has pronounced wiggles on large scales for traditional reconstruction, but is close to zero in the field-level cases which remove RSD by design. %\aeb{show to higher k}
} 
\label{fig:bao_sn}
\end{figure*}

To further quantify the reconstruction quality, the top panel of Figure \ref{fig:bao_fit} shows the monopole and quadrupole of the reconstructed fields, as well as the model best fit, averaged over 900 realizations. The amplitude for traditional reconstruction is largest as traditional reconstruction does not undo galaxy biasing. On the other hand, the field-level approaches produce approximately unity bias fields and thus have comparable amplitude to one another. The quadrupole for both field-level cases is approximately zero, as, unlike traditional reconstruction, they are designed to construct the linear field, which is free from RSD. Explicit inference provides a biased estimate of the linear field, hence there is a small non-zero quadrupole, and the monopole reconstructs more power on large scales and less on small scales compared to the implicit approach. This is due to misspecification in the explicit model and likelihood for the scales being reconstructed. Nevertheless, the exact shape of the reconstructed power is not important for BAO reconstruction, as we will marginalize over the broadband features when fitting the BAO. The lower panels show that the fit residuals are all sub-percent, implying high accuracy.

The quality of BAO reconstruction is fundamentally related to the sharpness of the BAO wiggles in the reconstructed power. Figure \ref{fig:bao_sn} shows the signal-to-noise of the wiggle-component of the power spectrum, where the noise is taken as the Gaussian covariance. The top panel shows the monopole: on large scales, the wiggles produced by all three methods agree, while on small scales the sharpest wiggles are produced by implicit field-level inference, followed by explicit, and then traditional. This agrees with the intuition above, that a higher quality reconstruction on small scales produces sharper BAO wiggles. The lower panel shows the quadrupole, which has pronounced wiggles on large scales for traditional reconstruction, but is close to zero in the field-level cases which remove RSD by design.

\begin{figure*}[ht!]
\includegraphics[width=\textwidth]{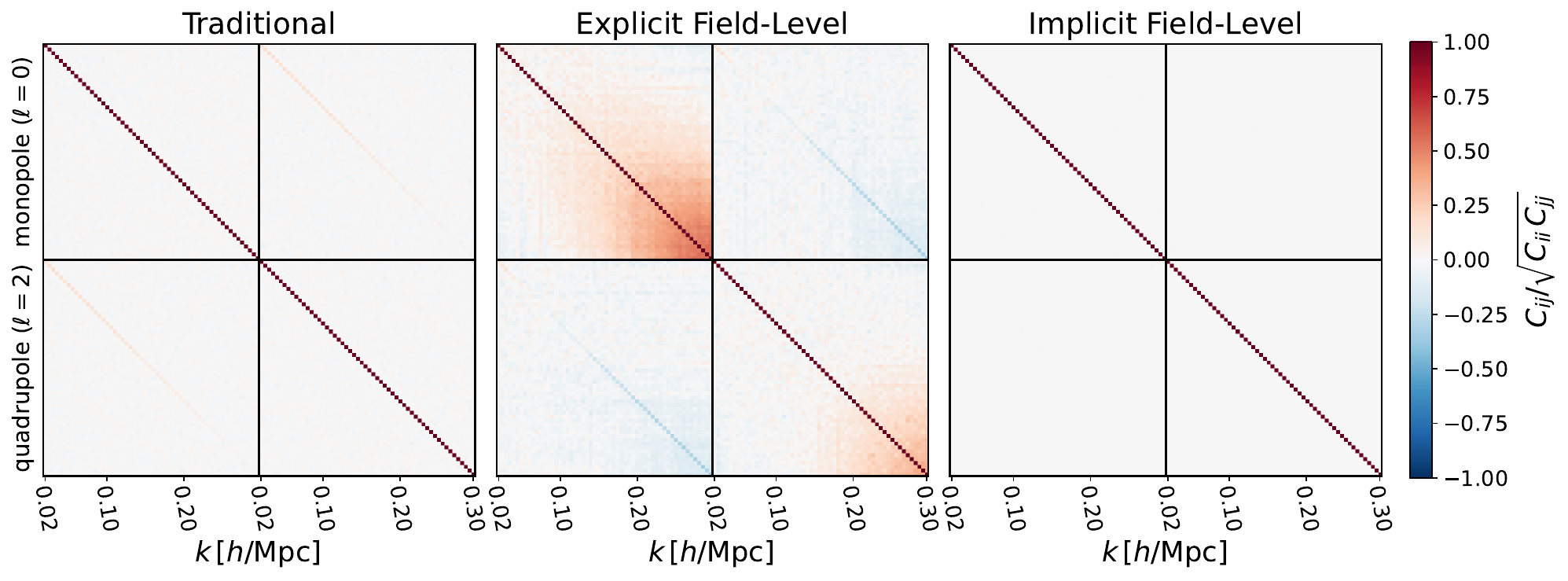}
\caption{
\textbf{Correlation matrix} for traditional (left), explicit field-level (center), and implicit field-level (right) reconstruction. Traditional shows a block-diagonal structure, explicit shows a more complex correlation structure, while implicit is diagonal and agrees with the theoretical Gaussian covariance at the percent level.} 
\label{fig:corr_mat}
\end{figure*}

\begin{figure}[ht!]
\hspace{-1.2em}
\includegraphics[width=0.5\textwidth]{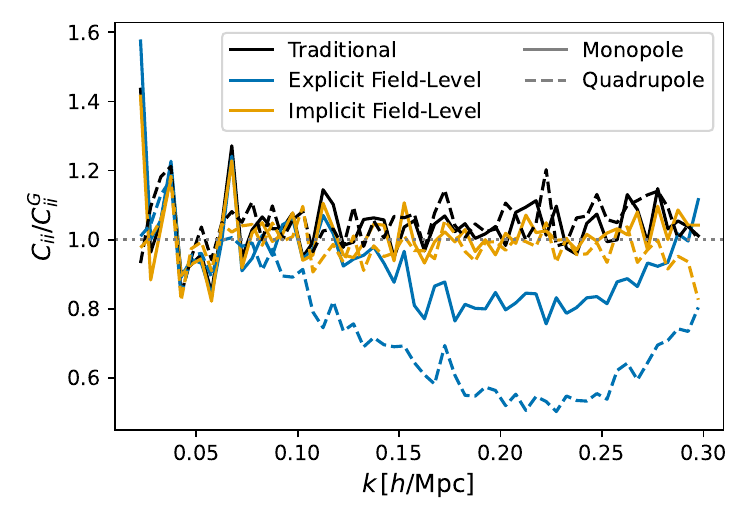}
\caption{
\textbf{Covariance matrix} ratio of diagonal components between numerical and Gaussian calculation for traditional (left), explicit field-level (center), and implicit field-level (right) reconstruction. Implicit shows the best agreement with the theoretical Gaussian covariance, closely followed by traditional reconstruction, while explicit does not due to misspecification of the likelihood.} 
\label{fig:cov_ratio}
\end{figure}

\begin{figure*}[ht!]
\includegraphics[width=0.5\textwidth]{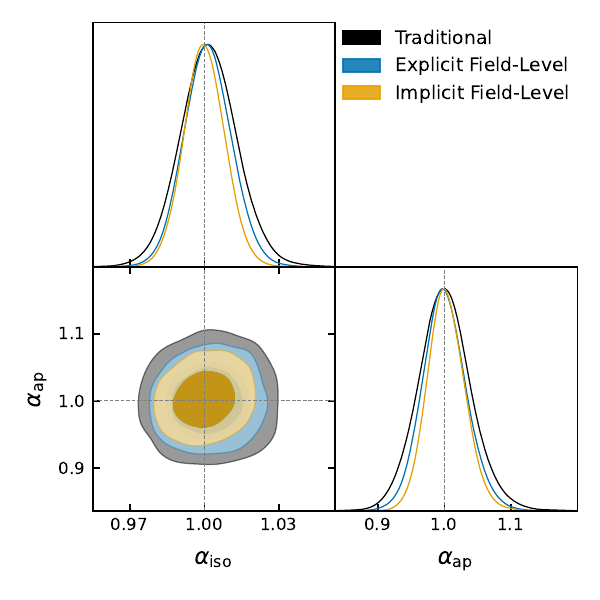}
\includegraphics[width=0.5\textwidth]{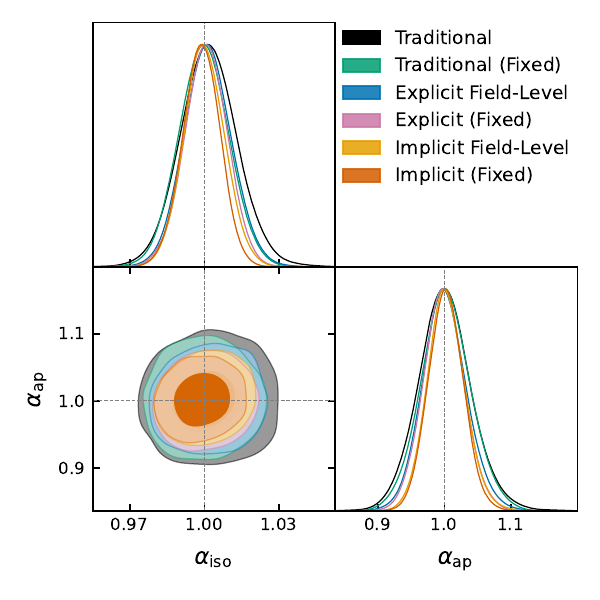}
\caption{
\textbf{BAO constraints} using uniform priors (left) and overlayed with fixed priors (right).  The inference is unbiased in all cases, giving $\alpha_{\rm iso}=\alpha_{\rm ap}=1$ (dotted lines). With uniform priors, there is a 29\% (17\%) improvement in constraining power for $\alpha_{\rm iso}$ for implicit (explicit) field-level reconstruction compared to traditional. 
%This corresponds to a two times improvement in contour area. 
When fixing the nuisance parameters, the constraining power improves for all methods, although we will see later that traditional reconstruction is not robust to model misspecification when fixing. When fixing, the improvement becomes 35\% (26\%) for implicit (explicit) field-level reconstruction compared to traditional with uniform priors. The improvements for $\alpha_{\rm ap}$ are similar, as reported in Table \ref{tab:summary}.
}
\label{fig:bao_tri}
\end{figure*}

\begin{figure*}[ht!]
\includegraphics[width=\textwidth]{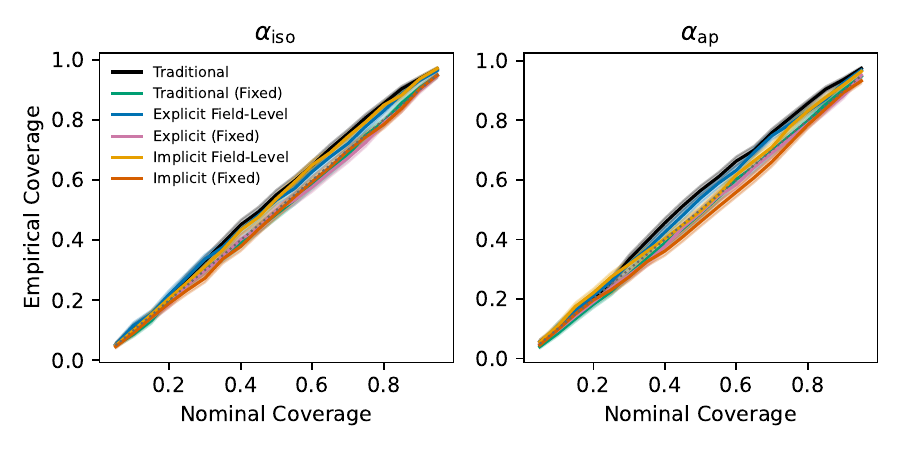}
\caption{
\textbf{Coverage test} for $\alpha_{\rm iso}$ (left) and $\alpha_{\rm ap}$ (right). In the case of uniform priors (black, blue, gold) there is a slight underconfidence, which is improved by using fixed nuisance parameters (green, pink, red).
} 
\label{fig:bao_cov}
\end{figure*}

To gain intuition for the information content, Figure \ref{fig:corr_mat} shows the numerical correlation matrices for the three reconstruction methods. Traditional reconstruction is block diagonal, with positive correlation within and between the monopole and quadrupole. Explicit field-level inference has a more complex correlation structure (particularly on small scales), with negative correlations between the monopole and quadrupole, and between high and low scales in the monopole---this occurs due to misspecification in the explicit model and likelihood for the scales being reconstructed and/or because of bias in the MAP estimate. On the other hand, implicit field-level inference is diagonal. Additionally, Figure \ref{fig:cov_ratio} shows the ratio of the diagonal components of the covariance with respect to the theoretical Gaussian covariance of equation (\ref{eqn:gauss_cov}). Both traditional and implicit field-level inference show good agreement, implying that implicit inference is accurately reconstructing the linear density field, and that simply using the Gaussian covariance is accurate.

We now explicitly study the information content by fitting the BAO to the reconstructed monopole and quadrupole. Figure \ref{fig:bao_tri} (left) shows constraints on the BAO scale parameters for the different reconstruction methods using uniform or fixed priors for the nuisance parameters. The data vector is the monopole and quadrupole averaged over 900 realizations. In the case of uniform priors, we marginalize over the broadband parameters analytically, and the other nuisance parameters numerically---we show the full triangle contour plot in Appendix \ref{app:marg} Figure \ref{fig:bao_tri_all}. 
Table \ref{tab:summary} shows the inference results, which are unbiased in all cases, giving $\alpha_{\rm iso}=\alpha_{\rm ap}=1$. For $\alpha_{\rm iso}$, with uniform priors there is a 29\% (17\%) improvement in constraining power for implicit (explicit) field-level reconstruction compared to traditional. When fixing the nuisance parameters to the MAP values (given in Appendix \ref{app:marg} Table \ref{tab:map}), the improvement increases to 35\% (26\%). For $\alpha_{\rm ap}$, the corresponding improvements are 28\% (15\%) with uniform priors and 35\% (25\%) when fixing. Combining both parameters, the figure of merit improves by a factor of $2.0\times$ ($1.4\times$) for implicit (explicit) with uniform priors, and $2.4\times$ ($1.8\times$) when fixing the nuisance parameters. 
We theoretically interpret the source of the increased information in terms of the wiggle and no-wiggle component in Appendix \ref{app:newfit}.

Having demonstrated unbiased inference, we now perform coverage tests to validate the quoted improvement in constraining power and to provide some insight on what priors to choose. To our knowledge this is the first coverage test of traditional and field-level reconstruction. Unlike in the previous discussion where we averaged the data-vector over 900 realizations before performing the fit, we now perform the fit on each of the 900 realizations individually and then ask how often the truth lies within the confidence interval around the inferred value---i.e.~how the empirical coverage compares to the nominal coverage. 
Figure \ref{fig:bao_cov} shows coverage plots for the different reconstructions when using uniform or fixed priors on the nuisance parameters. The diagonal dotted line corresponds to perfect coverage---anything above implies underconfidence (error bars too large), while anything below implies overconfidence (error bars too small). In all cases the line is almost diagonal, however, the uniform prior lines lie slightly above the diagonal, implying underconfidence. On the other hand, fixing the nuisance parameters to their MAP values, shows almost perfect coverage in all cases (while implicit field-level becomes slightly overconfident for $\alpha_{\rm ap}$, in practice we are only concerned about the 1-sigma error bar and beyond, i.e.~$>68\%$, where the coverage follows the diagonal---moreover, we find that the RMSE matches the average standard deviation). This implies that using tight, simulation-informed priors is both more informative and well covered for 
explicit and implicit field-level inference.

\begin{figure*}[t!]
\includegraphics[width=\textwidth]{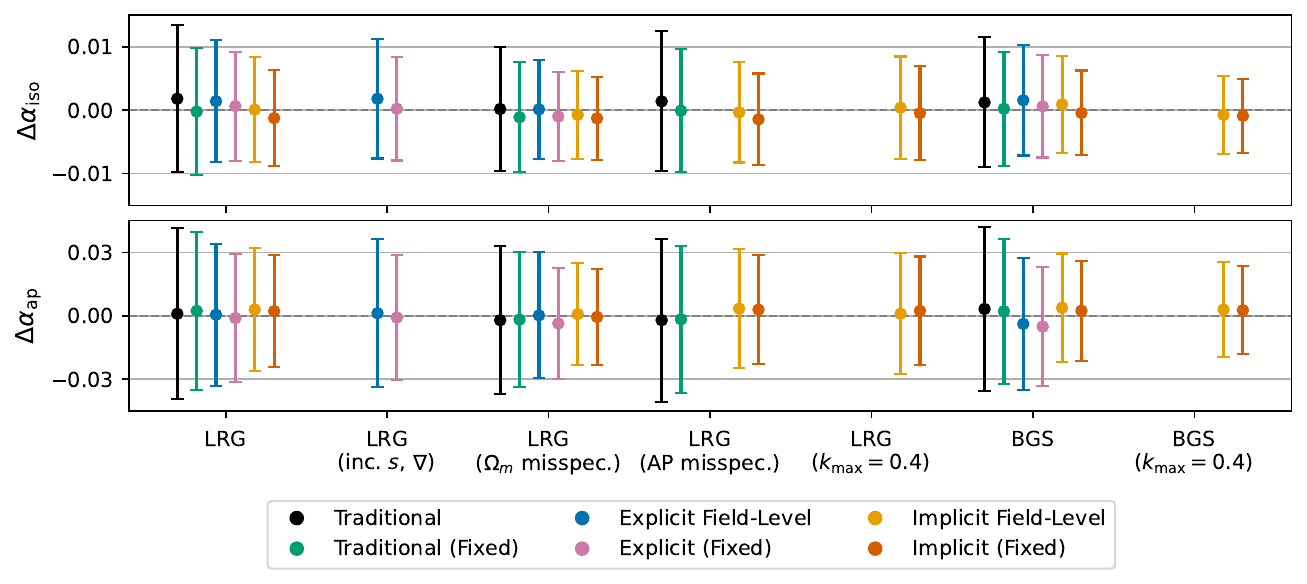}
\caption{
\textbf{Summary} of all power-spectrum extracted BAO constraints discussed throughout this work. All values correspond to the difference between the inferred value and the true values, $\Delta \alpha$. Different colors signify different reconstruction algorithms. Different batches of constraints correspond to, from left to right, the fiducial LRG analysis with $k_{\rm max}=0.3\,h/{\rm Mpc}$, LRG analysis including $s^2$ and $\nabla^2$ terms for explicit inference, LRG analysis with misspecified $\Omega_m$, LRG analysis with misspecified AP distortion, LRG analysis with $k_{\rm max}=0.4\,h/{\rm Mpc}$, fiducial BGS analysis with $k_{\rm max}=0.3\,h/{\rm Mpc}$, and BGS analysis with $k_{\rm max}=0.4\,h/{\rm Mpc}$. Key results are the improvement in constraining power when using field-level inference, the further improvement when pushing to smaller scales ($k_{\rm max}=0.4\,h/{\rm Mpc}$) for BGS, and the robustness to different types of model misspecification.
}
\label{fig:summary}
\end{figure*}

\subsection{Model Misspecification} \label{sec:results-misspec}

%Would be great to show for implicit (nw/*w) and explicit (nw/w) Onw plot with nw one plot with fli w. W

One of the key benefits of traditional BAO reconstruction is its robustness to model misspecification, particularly with respect to the fiducial cosmology \citep{cosmology_dependence}. We now explore how this translates to the field-level context. We first investigate the impact of an incorrect fiducial $\Omega_m$ parameter to test different sources of misspecification (see Section \ref{sec:robust}). 
Figure \ref{fig:summary} shows unbiased constraints for all reconstruction methods, implying accuracy in the face of model misspecification. We discuss extra results in Appendix \ref{app:add}: Figure \ref{fig:bao_tri_extra} (left) shows the contour plot and Figure \ref{fig:bao_cov_extra} (top) shows the corresponding coverage plot, where equivalent bias and coverage is achieved as in the case without model misspecification, except in the case of traditional reconstruction with fixed priors where there is slight overconfidence.

Another key source of model misspecification is an incorrect conversion of observed distances to comoving distances. As described in Section \ref{sec:robust}, we perform an AP test, by multiplying all $z$ coordinates by $3.9\%$, modifying both $\alpha_{\rm iso}$ and $\alpha_{\rm ap}$. 
Figure \ref{fig:summary} shows unbiased constraints for all reconstruction methods.
Figures \ref{fig:bao_tri_extra} (center) and \ref{fig:bao_cov_extra} (center) show constraints are unbiased and well covered in all cases. We do not run the AP test for explicit field-level as cuboid geometries are not currently supported in \texttt{pmwd}.

\subsection{BGS} \label{sec:results-bgs}

\begin{figure*}[ht!]
\includegraphics[width=\textwidth]{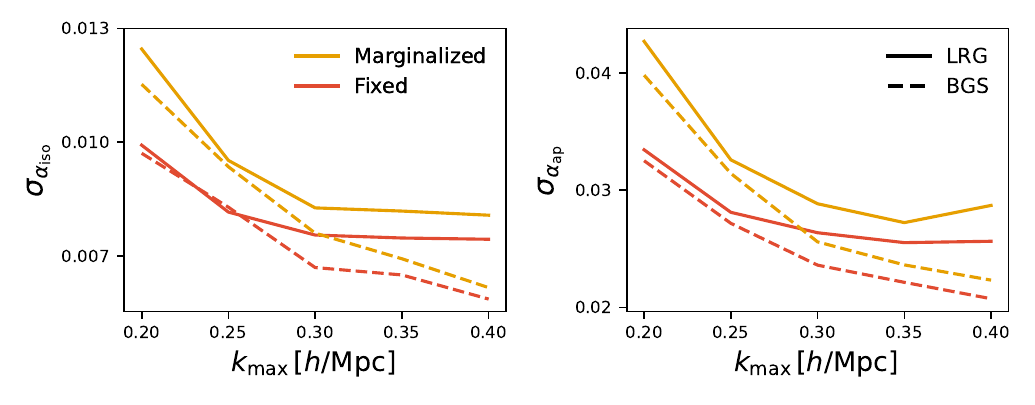}
\caption{
\textbf{BAO constraints as a function of $k_{\rm max}$} for LRG (solid) and BGS (dashed) with uniform (yellow) and fixed (red) nuisance parameters. In the case of uniform priors, there is a 4\% improvement in constraining power when going to $k_{\rm max}=0.3$ to $0.4\,h/{\rm Mpc}$, while for the higher number density BGS sample there is a more significant 20\% improvement, resulting in a three-fold improvement in constraining power compared to traditional methods. In the case of fixed priors the corresponding improvement is  1\% and 12\%, respectively. 
}
\label{fig:kmax}
\end{figure*}

We now perform an identical analysis for BGS galaxies instead of LRGs. Figure \ref{fig:summary} and Table \ref{tab:summary} presents the constraints on the $\alpha$ parameters. All methods have tighter constraints compared to LRGs due to BGS's higher number density. 

For $\alpha_{\rm iso}$, with uniform priors there is a 25\% (15\%) improvement in constraining power for implicit (explicit) field-level reconstruction compared to traditional. When fixing the nuisance parameters, the improvement increases to 34\% (22\%). For $\alpha_{\rm ap}$, the corresponding improvements are 33\% (21\%) with uniform priors and 38\% (28\%) when fixing. Combining both parameters, the figure of merit improves by a factor of $2.0\times$ ($1.5\times$) for implicit (explicit) with uniform priors, and $2.5\times$ ($1.8\times$) when fixing the nuisance parameters. 
Interestingly, the relative improvement of field-level methods to the traditional method is similar for BGS and LRGs. While this may seem somewhat surprising given BGS's higher number density, we will see in the next subsection that additional information can be found at higher $k$.
Figure \ref{fig:bao_tri_extra} (right) shows the unbiased contours for BGS, and Figure \ref{fig:bao_cov_extra} (bottom) shows similarly good coverage results for BGS as for the LRGs.

\subsection{Wiggling more out of the small scales} \label{sec:results-kmax}

%\begin{figure}[ht!]
%\includegraphics[width=0.5\textwidth]{paper_plots/fig-nb/fig-smth.pdf}
%\caption{
%\textbf{Smoothing dependence of traditional.}} 
%\label{fig:smth}
%\end{figure}

While $k_{\rm max} = 0.3\,h/{\rm Mpc}$ is the nominal choice for traditional reconstruction, field-level reconstruction improves reconstruction quality on smaller scales, thus BAO constraints could be expected to improve by increasing $k_{\rm max}$. Figure \ref{fig:kmax} shows the constraints on the BAO scale parameters as a function of $k_{\rm max}$ for implicit field-level reconstruction. We do not complete the analysis for explicit field-level inference as we find it is model misspecified for $k_{\rm max}=0.4\,h/{\rm Mpc}$. When marginalizing over nuisance parameters with uniform priors, for LRGs there is a small 4\% improvement in constraints on $\alpha_{\rm iso}$ from going from $k_{\rm max}=0.3$ to $0.4\,h/{\rm Mpc}$, while $\alpha_{\rm ap}$ constraints get worse. This implies that for LRGs we have already saturated the information by $k_{\rm max}=0.3$ and are dominated by shot noise beyond.
On the other hand, for the higher number density BGS sample there is a more significant 20\% improvement for both $\alpha$ parameters. A more gentle improvement in constraining power occurs in all cases when nuisance parameters are fixed. 
Thus, for BGS with $k_{\rm max}=0.4\,h/{\rm Mpc}$, implicit field-level inference improves constraints on $\alpha_{\rm iso}$ by 42\% (39\%) and $\alpha_{\rm ap}$ by 46\% (44\%) when using fixed (uniform) priors, relative to traditional reconstruction with uniform priors. This corresponds to a $3.2\times$ improvement in FoM.
Hence, field-level reconstruction not only improves the BAO constraints for a fixed set of scales, but it also enables the extraction of information from smaller scales than is possible with traditional reconstruction. A higher density tracer, such as BGS, benefits more from this due to its lower shot noise. %Moreover, there could be further gains from going to even smaller scales.

\subsection{Summary and Extra Results} \label{sec:results-summary}

In addition to Figure \ref{fig:summary}, Table \ref{tab:summary} provides a summary of all the inferences performed and discussed in this work with correct fiducial cosmology. We additionally include a comparison of using the 2-point correlation function $\xi(s)$ instead of the power spectrum $P(k)$ in the lower rows using the same bins as in \cite{Paillas:2024cru}. We opted for $P(k)$ in the main text as it 
has a simpler covariance structure; however, using $\xi(s)$ could be beneficial in the presence of survey windows. We find that the overall trends for $\xi(s)$ are the same as $P(k)$, with implicit outperforming explicit outperforming traditional. There are some small changes in the constraints due to differences between working in configuration space and Fourier space, consistent with \cite{Chen:2024tfp}.

\begin{table*}[!th]
    \centering
    \small
    \begin{tabular}{l||cc|cc|cc}
        \hline
                     & \multicolumn{2}{c}{LRG} & \multicolumn{2}{c}{BGS} & \multicolumn{2}{c}{BGS $(k_{\rm max}=0.4\,h/{\rm Mpc})$} \\
        \hline\hline
        Method & $\alpha_{\rm iso} \, (\mu\pm\sigma)$ & $\alpha_{\rm ap} \, (\mu\pm\sigma)$ & $\alpha_{\rm iso} \, (\mu\pm\sigma)$ & $\alpha_{\rm ap} \, (\mu\pm\sigma)$ & $\alpha_{\rm iso} \, (\mu\pm\sigma)$ & $\alpha_{\rm ap} \, (\mu\pm\sigma)$ \\
        \hline
        \textbf{Uniform ($P$)} & & & & & & \\
        Traditional           & $1.0018 \pm 0.0116$ & $1.001 \pm 0.040$ & $1.0012 \pm 0.0102$ & $1.003 \pm 0.039$ & -- & -- \\[2pt]
        Explicit Field-Level  & $1.0014 \pm 0.0096$ & $1.001 \pm 0.034$ & $1.0016 \pm 0.0087$ & $0.996 \pm 0.031$ & -- & -- \\
        \quad\scriptsize\color{gray}{\% improvement}  & \scriptsize\color{gray}{17\%} & \scriptsize\color{gray}{15\%} & \scriptsize\color{gray}{15\%} & \scriptsize\color{gray}{21\%} & & \\[2pt]
        Implicit Field-Level  & $1.0001 \pm 0.0082$ & $1.003 \pm 0.029$ & $1.0009 \pm 0.0076$ & $1.004 \pm 0.026$ & $0.9993 \pm 0.0062$ & $1.003 \pm 0.022$ \\
        \quad\scriptsize\color{gray}{\% improvement}  & \scriptsize\color{gray}{29\%} & \scriptsize\color{gray}{28\%} & \scriptsize\color{gray}{25\%} & \scriptsize\color{gray}{33\%} & \scriptsize\color{gray}{39\%} & \scriptsize\color{gray}{44\%} \\
        \hline
        \textbf{Fixed} ($P$) & & & & & & \\
        Traditional           & $0.9998 \pm 0.0100$ & $1.002 \pm 0.037$ & $1.0002 \pm 0.0090$ & $1.002 \pm 0.034$ & -- & -- \\
        \quad\scriptsize\color{gray}{\% improvement}  & \scriptsize\color{gray}{14\%} & \scriptsize\color{gray}{8\%} & \scriptsize\color{gray}{12\%} & \scriptsize\color{gray}{13\%} & & \\[2pt]
        Explicit Field-Level  & $1.0006 \pm 0.0086$ & $0.999 \pm 0.030$ & $1.0006 \pm 0.0080$ & $0.995 \pm 0.028$ & -- & -- \\
        \quad\scriptsize\color{gray}{\% improvement}  & \scriptsize\color{gray}{26\%} & \scriptsize\color{gray}{25\%} & \scriptsize\color{gray}{22\%} & \scriptsize\color{gray}{28\%} & & \\[2pt]
        Implicit Field-Level  & $0.9988 \pm 0.0075$ & $1.002 \pm 0.026$ & $0.9996 \pm 0.0067$ & $1.002 \pm 0.024$ & $0.9991 \pm 0.0059$ & $1.003 \pm 0.021$ \\
        \quad\scriptsize\color{gray}{\% improvement}  & \scriptsize\color{gray}{35\%} & \scriptsize\color{gray}{35\%} & \scriptsize\color{gray}{34\%} & \scriptsize\color{gray}{38\%} & \scriptsize\color{gray}{42\%} & \scriptsize\color{gray}{46\%} \\
        \hline\hline
        \textbf{Uniform ($\xi$)} & & & & & & \\
        Traditional                & $1.0005 \pm 0.0124$ & $0.999 \pm 0.045$ & $1.0011 \pm 0.0108$ & $1.001 \pm 0.041$ & -- & -- \\
        Explicit Field-Level  & $1.0003 \pm 0.0108$ & $1.003 \pm 0.039$  & $0.9990 \pm 0.0082$ & $0.999 \pm 0.031$ & -- & -- \\
        \quad\scriptsize\color{gray}{\% improvement}  & \scriptsize\color{gray}{7\%} & \scriptsize\color{gray}{2\%} & \scriptsize\color{gray}{20\%} & \scriptsize\color{gray}{21\%} & & \\[2pt]
        Implicit Field-Level  & $0.9995 \pm 0.0083$ & $1.001 \pm 0.029$  & $1.0000 \pm 0.0071$ & $1.000 \pm 0.025$ & -- & -- \\
        \quad\scriptsize\color{gray}{\% improvement}  & \scriptsize\color{gray}{28\%} & \scriptsize\color{gray}{28\%} & \scriptsize\color{gray}{30\%} & \scriptsize\color{gray}{36\%} & & \\
        \hline
    \end{tabular}
    \caption{Summary of all non-model-misspecified BAO constraints discussed throughout this work. Gray percentages denote the reduction in $\sigma$ relative to Traditional reconstruction with uniform priors on $P(k)$; negative values indicate degradation.}
    \label{tab:summary}
\end{table*}

%Gaussian cov
%# Method	qiso	qap
%Traditional	1.001043 0.011366	1.001184 0.041106
%Explicit Field-Level	1.001811 0.011258	1.006060 0.043792
%Implicit Field-Level	1.000006 0.008303	1.003507 0.029475

% Another test -- coverage in 10-avgs. Maybe for later

\section{Conclusions} \label{sec:conclusions}

In this work, we presented the first application of field-level inference to reconstruct the BAO signal from galaxy mocks. Using DESI-like LRG and BGS mocks, we compared traditional reconstruction to two field-level approaches: explicit inference using a differentiable particle-mesh forward model with HEFT, and implicit inference using a CNN. We then analyzed the reconstructed fields with the DESI BAO fitting pipeline, enabling a direct comparison to standard survey methodology. Beyond demonstrating improved constraining power, we also carried out extensive tests of bias, coverage, and robustness.

Our analysis yielded several key findings:
\begin{itemize}
    \item \textbf{Significant Information Gain:} Both field-level methods outperform traditional reconstruction by recovering information from nonlinear scales. For LRGs, explicit inference improves constraints on the BAO scale parameters by 17--26\%, while implicit inference achieves gains of 29--35\%, depending on the choice of prior. For BGS, the improvement reaches 42-46\% in constraining power compared to traditional reconstruction, corresponding to \textit{a factor of 3.2 increase in FoM}.
    \item \textbf{Unlocking Small Scales:} The benefits of field-level reconstruction are further pronounced for high-density and low-redshift tracers. For the BGS sample, the implicit field-level reconstruction extracts information from smaller scales ($k_{\rm max} = 0.4\,h/\mathrm{Mpc}$) that are too small for traditional reconstruction to model accurately. 
    \item \textbf{Advantage of Implicit Inference:} We observe that the CNN-based implicit approach yields tighter constraints than the explicit approach. This suggests that the neural network successfully learns an effective likelihood that captures small-scale clustering and halo-galaxy connection physics more accurately than HEFT does in our explicit forward model.  %Other explanations may however still be possible, with explicit inference potentially improvable my sampling instead of optimizing. 
    \item \textbf{Simple Covariance:} We showed that the implicit field-level reconstruction produces a linear density field with a diagonal covariance matrix that matches the theoretical Gaussian prediction to percent-level accuracy. This suggests that complex, simulation-based covariance matrices may not be necessary for downstream analysis of implicit field-level reconstructed data.
    \item \textbf{Optimality:} Implicit field-level reconstruction yields nearly identical BAO constraints in Fourier space ($P(k)$) and configuration space ($\xi(s)$). This is expected: the reconstructed field has a diagonal, Gaussian covariance (Figure~\ref{fig:cov_ratio}), and for a Gaussian field the power spectrum is a sufficient statistic. This suggests the two-point function has saturated the available BAO information content.
    \item \textbf{Interpreting the Source of Information:} We demonstrate that the improvement in constraining power stems directly from the better restoration of the acoustic oscillation feature (wiggles) on small scales. By explicitly modeling the wiggle and no-wiggle components, we show that field-level reconstruction enhances the contrast of the wiggles and reduces their damping (Figure \ref{fig:bao_sn}). We motivate this theoretically in Appendix \ref{app:newfit}, also discussing the impact of the broadband (no-wiggle) component.
    \item \textbf{Robustness and Coverage:} Using 1,000 mock realizations, we performed extensive coverage tests for field-level BAO reconstruction. We demonstrated that all methods are unbiased in the presence of model misspecification (specifically, an incorrect fiducial $\Omega_m$ and incorrect AP rescaling). Varying $\Omega_m$ induces changes in bias, number density, and nonlinear tracer properties, and therefore provides a useful test of robustness to a broad class of modeling mismatches. We show our methods maintain accuracy and coverage when using tight, simulation-informed priors on nuisance parameters. %, whereas traditional reconstruction becomes overconfident in this regime.
\end{itemize}

Taken together, these results show that field-level reconstruction can deliver substantially tighter and statistically reliable BAO measurements in realistic galaxy samples, while remaining compatible with the traditional modular reconstruct-then-fit analysis pipeline.
Building on these encouraging results, future work will focus on applying these methods directly to survey data, accounting for observational systematics such as survey masking, selection effects, light cone effects, and fiber collisions. These systematics can be addressed by including such effects in the forward model for explicit field-level inference and in the training data for implicit field-level inference. Moreover, the subgrid approach used to patch the sky ensures the method naturally extends to reconstruct complex window geometries. Survey masking, selection effects, and light cone effects are predominantly large-scale and thus affect traditional and field-level methods similarly. Fiber collisions, which preferentially affect small scales, will need to be carefully forward modeled, although their impact will be naturally mitigated in future DESI data releases as the survey performs multiple passes and increases completeness.

Beyond constraining power, the two field-level approaches present complementary tradeoffs. Explicit inference requires no training data and can in principle be applied to any cosmology or tracer without retraining---making it naturally suited for joint inference of cosmological parameters and initial conditions. However, it requires a fast differentiable forward model, which often sacrifices simulation fidelity for differentiability (e.g.~fewer particles and simplified matter--galaxy connection), is computationally expensive per realization, and is limited by the assumed form of the likelihood. Implicit inference, by contrast, can be trained on any simulator at full fidelity, is fast at inference time, and learns an effective likelihood from simulations. Its main limitation is the dependence on training data: changes to the tracer, cosmology, or observational systematics require retraining. These tradeoffs suggest that both approaches have a role to play in the future of field-level inference.

There are also various avenues to utilize these results for even greater gains in cosmological parameter inference. First, improvements in signal-to-noise may be achievable by incorporating mass weighting into the reconstruction pipeline  \citep{Parker:2025mtg}. Second, while we only considered individual tracers in this analysis, one can perform an optimal combined field-level analysis of all the DESI tracers \citep{Valcin_DESI:2025yuk}, and ultimately multi-probe analyses by combining with weak lensing, the CMB, and more, using correlated simulations \citep{HalfDome}.
Third, while we utilized optimization to find the MAP estimate for explicit inference, a fully Bayesian analysis using Monte Carlo sampling \citep{Bayer:2023rmj, Simon:2025gwa} could properly propagate uncertainties in the reconstructed field, provided the posterior is well defined---this could be explored using the reconstruct-then-fit approach, as in this work, or jointly fitting for the BAO scale while performing reconstruction \citep{Babic:2022dws, Babic:2024wph}. Fourth, improvements to the forward model, such as higher-order bias expansions, emulators \citep{modi2018cosmological, CosmicRIM, Jamieson:2022lqc, Jamieson:2024fsp, doeser2024bayesian, doeser2025learning, CHARM}, or differentiable halo and galaxy modeling \citep{jfof, diffhod}, could reduce the small-scale modeling error in the explicit approach. Finally, it would be fruitful future work to compare with other machine learning methods to reconstruct the initial conditions, such as optimal transport \citep{Nikakhtar:2022cik, Nikakhtar:2023ohj}.

Although this work focused on the BAO feature, as it is a robust standard ruler and thus a promising first approach of field-level reconstruction to spectroscopic survey data, the reconstructed initial conditions contain a wealth of additional information which we will explore in future works. These fields can, for example, be used to reconstruct velocities \citep{Bayer:2022vid}, constrain RSD, and primordial non-Gaussianity \citep{Chen:2024exy, floss2024improving, Bottema:2025vww} with higher fidelity than conventional approaches. Ultimately, our field-level inference pipelines can be used to perform joint inference of the initial field and cosmological parameters. Our results demonstrate that field-level inference is a powerful and viable  tool for maximizing the scientific return of current and future spectroscopic surveys.

\section*{Acknowledgments}
We thank Kazuyuki Akitsu, Shadab Alam, Raul Angulo, Ivana Babić, Xinyi Chen, Carolina Cuesta-Lazaro, Arnaud de Mattia, Natalí de Santi, Zhejie Ding, Chenze Dong, Daniel Eisenstein, Richard Feder, Thomas Flöss, Shirley Ho, Benjamin Horowitz, François Lanusse, Yin Li, Avi Loeb, Patrick McDonald, Sheena Meng, Seshadri Nadathur, Nhat-Minh Nguyen, Nikhil Padmanabhan, Enrique Paillas, Will Percival, Oliver Philcox, Michael Rashkovetskyi, Fabian Schmidt, Hee-Jong Seo, Blake Sherwin, David Spergel, Beatriz Tucci, Francisco Villaescusa-Navarro, Zvonimir Vlah, Martin White, and Nate for their insightful discussions and valuable assistance during the course of this work. 
The computations reported in this paper were performed using resources made available by the Flatiron Institute.
This research used resources of the National Energy Research Scientific Computing Center (NERSC), a Department of Energy Office of Science User Facility using NERSC award ASCR-ERCAP0029232. Support for this work was provided by NASA through the NASA Hubble Fellowship grant HST-HF2-51572.001 awarded by the Space Telescope Science Institute,
which is operated by the Association of Universities for Research in Astronomy, Inc., for
NASA, under contract NAS5-26555.
This work is supported by NSF CDSE grant number AST-2408026 and NASA TCAN grant number 80NSSC24K0101. L.P. is supported by the NSF Graduate Research Fellowship.

\section*{Software}

This work made use of the \texttt{FastPM} \citep{Feng2016, Bayer_2021_fastpm} and \texttt{pmwd} \citep{li2022pmwd} packages for particle mesh simulations and differentiable forward modeling. 
Galaxies were painted and grid operations performed using \texttt{nbodykit} \citep{Hand_2018}. 
Traditional reconstruction was performed using \texttt{pyrecon} (\url{https://github.com/cosmodesi/pyrecon}). 
Boltzmann solver calculations were performed using \texttt{CLASS} \citep{CLASS}. 
Parameter inference and BAO fitting were conducted using \texttt{desilike} (\url{https://github.com/cosmodesi/desilike}), utilizing the \texttt{emcee} \citep{Foreman_Mackey_2013} sampler. 
We also acknowledge the use of \texttt{JAX} \citep{jax2018github} for automatic differentiation and high-performance numerical computing. 
The figures in this work were created using \texttt{Matplotlib} \citep{Hunter:2007}.
General numerical analysis relied on \texttt{NumPy} \citep{harris2020array} and \texttt{SciPy} \citep{virtanen2020scipy}.

\appendix

\renewcommand{\thefigure}{A\arabic{figure}}
\setcounter{figure}{0}
\section{Nuisance Parameters}
\label{app:marg}

\begin{figure}[ht!]
\includegraphics[width=\textwidth]{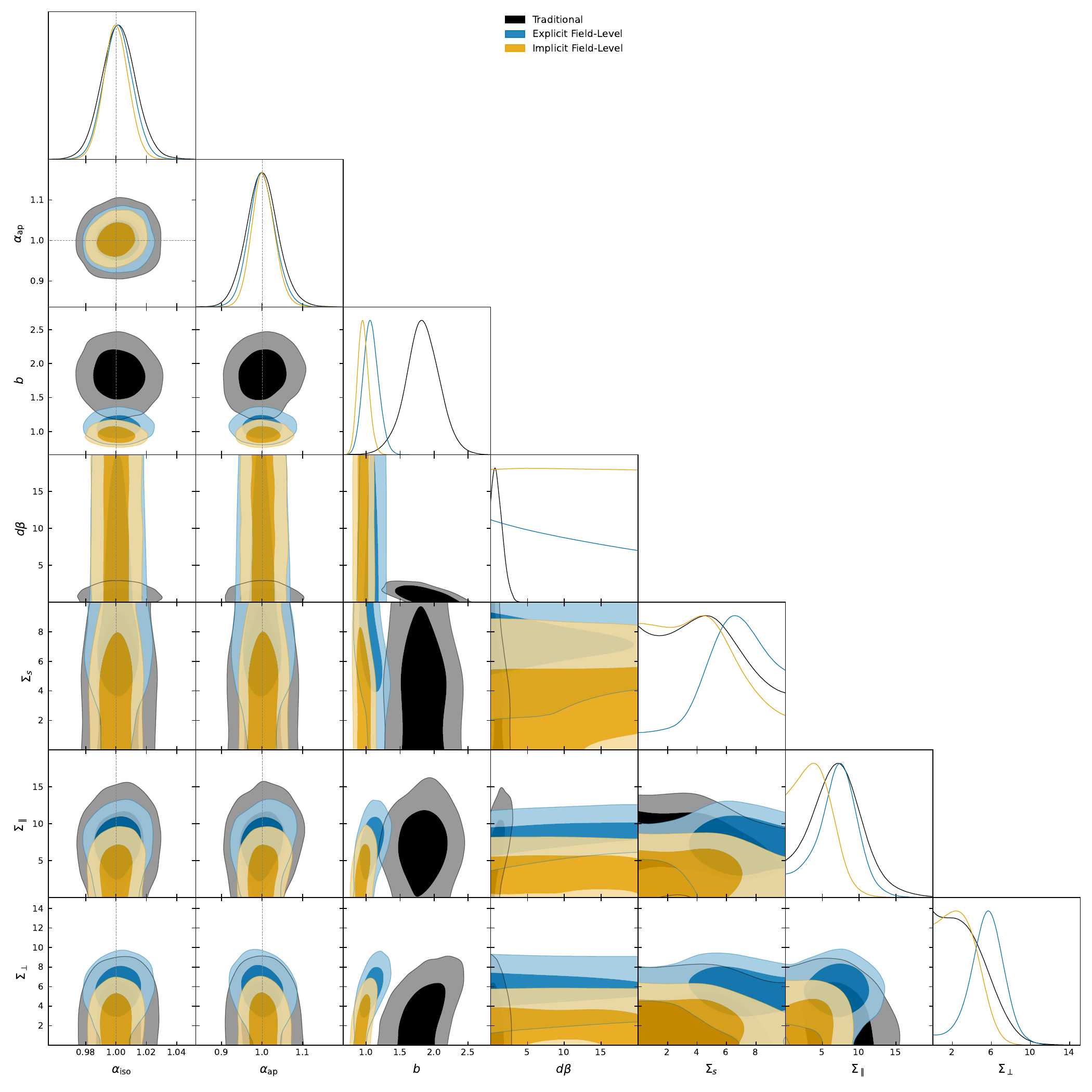}
\caption{
\textbf{BAO constraints with nuisance parameters.} Like Figure \ref{fig:bao_tri} (left), but with nuisance parameter constraints.} 
\label{fig:bao_tri_all}
\end{figure}

Figure \ref{fig:bao_tri_all} shows the full contour plot associated with Figure \ref{fig:bao_tri} (left), showing the posteriors of the nuisance parameters when uniform priors are used. Note we analytically marginalize over the $a_n$ parameters. Field-level approaches have poorly constrained $d\beta = f/f_{\rm fid}$, as expected as they are designed to remove RSD, corresponding to no Kaiser effect in the reconstructed field. In principle, one could fix $d\beta=0$ for field-level approaches, but in this work we preferred to let the model learn it itself. 

We report the best fit (MAP) values of the nuisance parameters for this case in Table \ref{tab:map}. These are used for the fixed prior results.
It can be seen that the field-level methods reduce the bias to approximately unity, as they predict the linear field.
Implicit field-level has approximately equal to or lower MAP values for the $\Sigma$ parameters, implying a smaller smoothing, as it most precisely reconstructs the smallest scales. The same is true for explicit field-level, except that it has a larger $\Sigma_\perp$, perhaps implying misspecification while modeling RSD on small scales with HEFT.
%BGS shows similar behavior, although in this case explicit field-level has a larger $\Sigma_\parallel$ rather than $\Sigma_\perp$.

\begin{table}[!th]
    \centering
    \small
    \begin{tabular}{l|ccccc}
        \hline
        Method & $b$ & $d\beta$ & $\Sigma_s$ & $\Sigma_\parallel$ & $\Sigma_\perp$ \\
        \hline\hline
        Traditional          & 1.83 & 0.94    & 5.36 & 6.61 & 2.60 \\
        Explicit Field-Level & 1.10 & -- & 5.07 & 4.92 & 5.66 \\
        Implicit Field-Level & 0.96 & -- & 4.47 & 3.90 & 2.75 \\
        \hline
    \end{tabular}
    \caption{Fixed nuisance parameter values (MAP estimates) for LRGs used in the fixed-prior fitting configuration, obtained by fitting the mean data vector over 900 realizations. All $\Sigma$ parameters are in units of ${\rm Mpc}/h$. We do not report $d\beta$ for the field level cases as they remove RSD by design and $d\beta$ is unconstrained.}
    \label{tab:map}
\end{table}

\renewcommand{\thefigure}{B\arabic{figure}}
\setcounter{figure}{0}
\section{Interpreting the source of information}
\label{app:newfit}

%To understand the source of the BAO information,
%we interpret the quality of reconstruction in terms of the
%cross-correlation coefficient. 
%This is naturally connected to the formalism for traditional reconstruction
%(e.g.~\cite{Chen:2024tfp}), outlined in Section~\ref{sec:fitting}.

To understand the source of the BAO information,
we work within the formalism for the traditional BAO fit
(e.g.~\cite{Chen:2024tfp}), outlined in Section~\ref{sec:fitting}.
We use an $RR$ superscript, e.g.~$P^{RR}(k,\mu)$, to denote the power spectrum of the
reconstructed field $\delta^R$. We denote by $P^{RR}_{\rm nw}(k,\mu)$ the corresponding reconstructed power spectrum obtained from initial conditions with the BAO wiggles removed.
Following Eq.~(\ref{eq:generic_model_chen}), we write the reconstructed power spectrum as
\begin{equation}
P^{RR}(k,\mu;\alpha)
=
B(k,\mu)\,P_{\rm nw}(k)
+
C(k,\mu)\,P_{\rm w}(k;\alpha)
+
D(k),
\label{eq:B1}
\end{equation}
where $P_{\rm nw}(k)$ is the linear no-wiggle power spectrum, $P_{\rm w}(k;\alpha)$ is the oscillatory wiggle component containing the BAO-scale dependence, and $D(k)$ is the residual broadband contribution. The corresponding no-wiggle reconstruction is
\begin{equation}
P^{RR}_{\rm nw}(k,\mu)
=
B(k,\mu)\,P_{\rm nw}(k)+D(k).
\end{equation}
The total linear power spectrum is
\begin{equation}
P(k)=P_{\rm nw}(k)+P_{\rm w}(k)\simeq P_{\rm nw}(k),
\end{equation}
where the final approximation reflects that the BAO wiggles are a small modulation of the broadband spectrum.

Approximating the covariance by the disconnected Gaussian term, the chi-squared may be written as
\begin{equation}
\chi^2(\alpha)
=
\sum_{k,\mu}
\frac{
\left[
P^{RR}_{\rm obs}(k,\mu)-P^{RR}(k,\mu;\alpha)
\right]^2
}{
2\left[P^{RR}_{\rm nw}(k,\mu)\right]^2
}.
\label{eq:B7}
\end{equation}
The use of $P^{RR}_{\rm nw}$ in the denominator reflects the approximation that the variance is controlled by the smooth reconstructed power, while the dependence on $\alpha$ enters through the oscillatory part. 
%In deriving the Fisher information below, we also neglect the $\alpha$-dependence of the covariance. 
The resulting expressions should therefore be interpreted as an approximate description of how the BAO information enters through the mean model, rather than as an exact decomposition of the full likelihood.

The Fisher information is then given by
\begin{equation}
F_{\alpha\alpha}
=
\frac{1}{2}
\left.
\frac{\partial^2 \chi^2}{\partial \alpha^2}
\right|_{\alpha=\alpha_{\rm fid}}
=
\frac12
\sum_{k,\mu}
\frac{
\left[
\partial_\alpha P^{RR}(k,\mu;\alpha)
\right]^2
}{
\left[P^{RR}_{\rm nw}(k,\mu)\right]^2
}.
\label{eq:B8}
\end{equation}
Since the BAO-scale dependence enters only through the wiggle term in Eq.~(\ref{eq:B1}), we have
\begin{equation}
\partial_\alpha P^{RR}(k,\mu;\alpha)
=
C(k,\mu)\,
\frac{\partial P_{\rm w}(k;\alpha)}{\partial \alpha},
\label{eq:B9}
\end{equation}
and therefore
\begin{equation}
F_{\alpha\alpha}
=
\frac12
\sum_{k,\mu}
\left[
\frac{C(k,\mu)}{P^{RR}_{\rm nw}(k,\mu)}
\frac{\partial P_{\rm w}(k;\alpha)}{\partial \alpha}
\right]^2.
\label{eq:B10}
\end{equation}
This shows that the BAO information is controlled by the wiggle response $\partial P_{\rm w}/\partial\alpha$, the wiggle transfer coefficient $C(k,\mu)$, and the smooth reconstructed power entering the denominator. Increasing the wiggle response or the transfer coefficient enhances the recovered oscillatory signal and therefore increases the BAO information, while increasing the no-wiggle reconstructed power raises the variance, making the wiggles and BAO information harder to resolve. More precisely, the relevant quantity is the ratio $C(k,\mu)/P^{RR}_{\rm nw}(k,\mu)$: a larger smooth reconstructed power does not necessarily imply less BAO information if it is accompanied by a corresponding increase in the transfer coefficient.

We now seek to interpret the role of the correlation coefficient $r$ on the BAO information. To do so, we work in the limit that the transfer varies smoothly with scale and does not strongly distinguish between the wiggle and no-wiggle components. This is not exact 
%is expected to be most accurate when the reconstruction response varies smoothly with scale; it will become less accurate 
in the presence of nonlinear mode coupling or reconstruction-induced effects imprinting additional damping on the wiggle component. In terms of the cross-spectrum between the reconstructed field and the no-wiggle initial linear field $\delta^I$, we therefore write
\begin{equation}
C(k,\mu)
\simeq
\left[\frac{P^{RI}_{\rm nw}(k,\mu)}{P_{\rm nw}(k)}\right]^2,
\label{eq:B3}
\end{equation}
and define the no-wiggle cross-correlation coefficient
\begin{equation}
r^2(k,\mu)
=
\frac{\left[P^{RI}_{\rm nw}(k,\mu)\right]^2}
{P^{RR}_{\rm nw}(k,\mu)\,P_{\rm nw}(k)}.
\label{eq:B4}
\end{equation}
This quantity measures the fraction of the reconstructed no-wiggle power that remains correlated with the linear no-wiggle field, and therefore provides a useful measure of reconstruction quality: values of $r\simeq 1$ indicate that most of the reconstructed power remains linearly correlated with the initial field, while smaller values indicate a larger residual stochastic or broadband contribution. In practice, the full cross-correlation coefficient constructed from the total (wiggle plus no-wiggle) power contains small residual oscillatory features. The approximation adopted here is that the dominant scale dependence relevant for the BAO response is captured by the no-wiggle quantity in Eq.~(\ref{eq:B4}).

Combining Eqs.~(\ref{eq:B3}) and (\ref{eq:B4}), we obtain
\begin{equation}
\frac{C(k,\mu)}{P^{RR}_{\rm nw}(k,\mu)}
=
\frac{r^2(k,\mu)}{P_{\rm nw}(k)}.
\label{eq:B5}
\end{equation}
This may also be written as
\begin{equation}
r^2(k,\mu)
=
\frac{C(k,\mu)\,P_{\rm nw}(k)}
{B(k,\mu)\,P_{\rm nw}(k)+D(k)}.
\label{eq:B6}
\end{equation}
This form is useful for interpretation: the numerator represents the BAO-bearing correlated contribution to the reconstructed field, while the denominator is the total smooth reconstructed power in the template. Residual broadband power that is not correlated with the linear field therefore reduces $r(k,\mu)$.

Eq.~(\ref{eq:B5}) allows Eq.~(\ref{eq:B10}) to be rewritten as
\begin{equation}
F_{\alpha\alpha}
=
\frac12
\sum_{k,\mu}
r^4(k,\mu)
\left[
\frac{1}{P_{\rm nw}(k)}
\frac{\partial P_{\rm w}(k;\alpha)}{\partial \alpha}
\right]^2,
\label{eq:B12}
\end{equation}
as in \cite{Parker:2025mtg}. In this limit, the BAO information is controlled by
(i) the intrinsic wiggle sensitivity, encoded by the fractional wiggle response $P_{\rm nw}^{-1}\partial P_{\rm w}/\partial\alpha\simeq d\ln P/d\alpha$, and
(ii) the reconstruction correlation, encoded by $r^4(k,\mu)$.
The reconstructed power spectrum $P^{RR}(k,\mu)$ does not enter independently, but only through its contribution to $r(k,\mu)$. %Consequently, a smaller reconstructed broadband power would not by itself guarantee improved BAO information. 

The equations outlined in this appendix also suggest a practical strategy for setting priors. The transfer coefficient $C(k,\mu)$ cannot be well determined from the data alone, since it is partially degenerate with the broadband terms in Eq.~(\ref{eq:generic_model_chen}). A natural approach is therefore to calibrate $C(k,\mu)$ from simulations using Eq.~(\ref{eq:B5}) and fit it to the ansatz of Eq.~(\ref{Chen3}), while allowing the remaining broadband contribution to be constrained directly from the data. In the main text we explore two limiting prior choices---uniform wide priors and fixed simulation-informed priors. More generally, one can learn the prior distribution from simulations, while fitting the broadband terms entering Eq.~(\ref{eq:generic_model_chen}) from the data, with the flexible contribution $D(k)$ parameterized as in Eq.~(\ref{eq:spline_basis}). We applied this approach for the analysis in this paper and found it gave larger values of $\Sigma_\parallel$ and $\Sigma_\perp$ than the full BAO fit MAP results (reported in Appendix \ref{app:marg} Table \ref{tab:map}), but that it nevertheless gave identical constraints on the $\alpha$ parameters. We leave a more thorough study of simulation-informed priors to future work.

\renewcommand{\thefigure}{C\arabic{figure}}
\setcounter{figure}{0}
\section{Additional contour and coverage plots}
\label{app:add}

\begin{figure}[t!]
\includegraphics[width=0.33\textwidth]{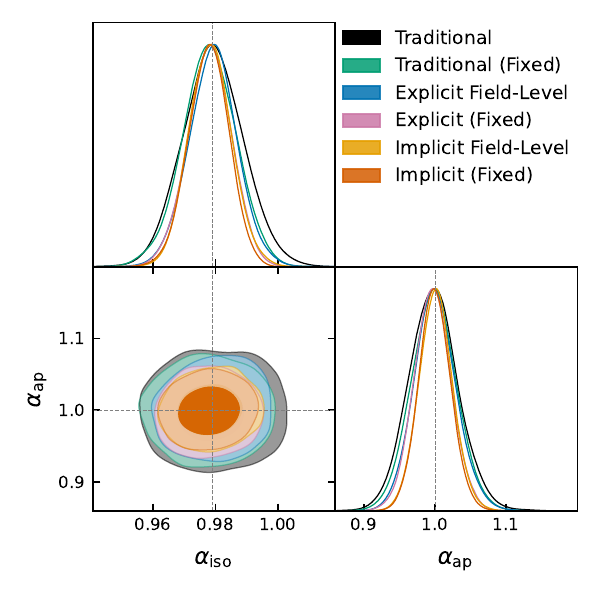}
\includegraphics[width=0.33\textwidth]{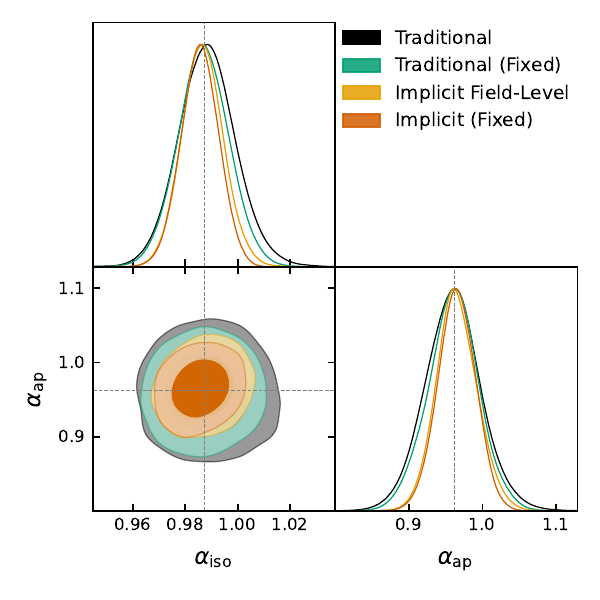}
\includegraphics[width=0.33\textwidth]{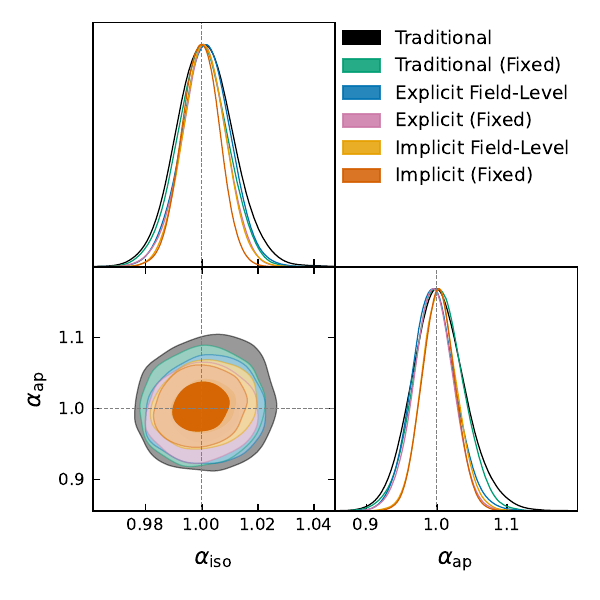}
\caption{
\textbf{BAO constraints}, like for the fiducial LRG plot of Figure \ref{fig:bao_tri}, but with misspecified $\Omega_m$ (left), misspecified AP distortion (center), and for BGS (right). All show unbiased constraints with similar trends to the fiducial LRG application. 
}
\label{fig:bao_tri_extra}
\end{figure}

\begin{figure*}[t!]
\centering
\includegraphics[width=0.8\textwidth]{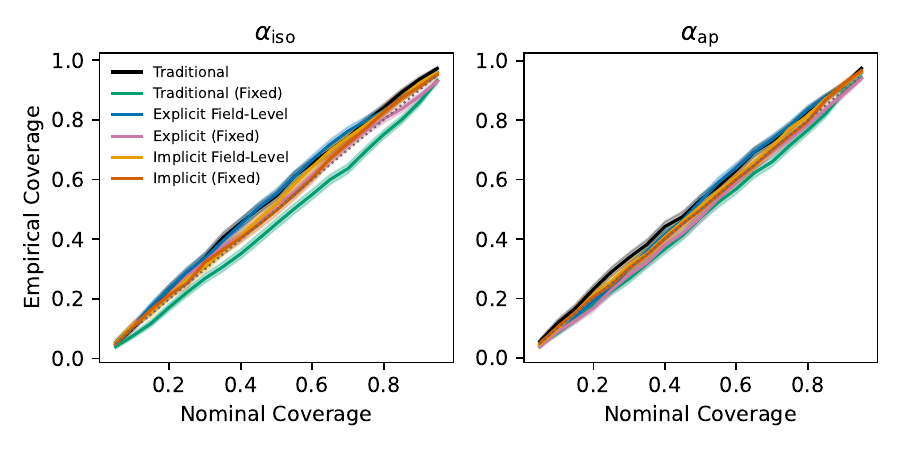} 
\includegraphics[width=0.8\textwidth]{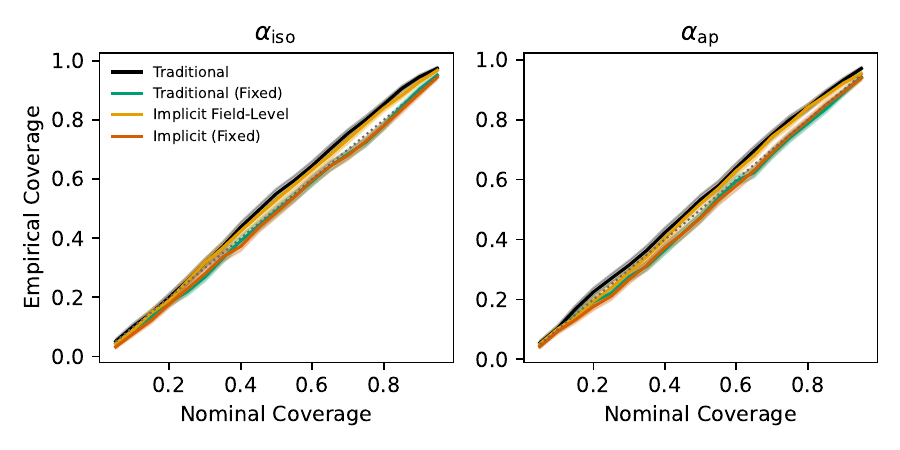} 
\includegraphics[width=0.8\textwidth]{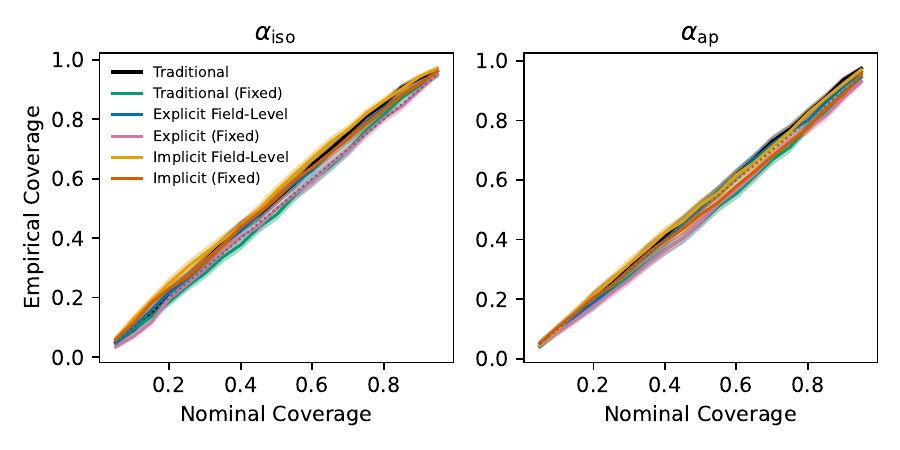} 
\caption{
\textbf{Coverage tests}, like for the fiducial LRG plot of Figure \ref{fig:bao_cov}, but with misspecified $\Omega_m$ (top), misspecified AP distortion (center), and for BGS (bottom). All methods are well covered, with the exception of traditional reconstruction with fixed nuisance parameters in the misspecified $\Omega_m$ case.
}
\label{fig:bao_cov_extra}
\end{figure*}

Figure \ref{fig:bao_tri_extra} shows the contour plot with misspecified $\Omega_m$ (left), misspecified AP distortion (center), and for BGS (right).  All show unbiased constraints with similar trends to the fiducial LRG application.  Figure \ref{fig:bao_cov_extra} shows the coverage plots for misspecified $\Omega_m$ (top), misspecified AP distortion (center), and for BGS (bottom). All methods are well covered, with the exception of traditional reconstruction with fixed nuisance parameters in the misspecified $\Omega_m$ case, which could be fixed by rescaling the covariance to the data as is often done in approaches such as \cite{Philcox2020:1904.11070}. 
This implies that care must be taken in choosing priors for nuisance parameters in traditional reconstruction, while fixing the nuisance parameters remains robust for field-level reconstruction. More generally, while we considered two extreme prior choices to build intuition---uniform and fixed---we advocate for learning the prior distribution from an ensemble of simulations to ensure ideal coverage.

\bibliography{refs}{}
\bibliographystyle{aasjournal}

\end{document}